\documentclass[a4paper]{article}

\usepackage{amsmath, amsthm, amsfonts, amssymb}

\usepackage{graphicx}
\graphicspath{{./Figures/}}
\usepackage{float}
\usepackage{multirow}
\usepackage{longtable}
\usepackage{subfig}
\usepackage{adjustbox}
\usepackage{array,makecell}
\usepackage{enumitem} 
\usepackage{mathtools}
\usepackage{mathdots}
\usepackage{changepage}
\usepackage{booktabs}
\usepackage{tcolorbox}
\usepackage{filecontents}
\usepackage{tikz}
\usepackage{tkz-graph}
\usepackage{chemfig}
\usepackage[title]{appendix}
\usepackage{pgfplots}
\pgfplotsset{compat=1.17}

\theoremstyle{definition}

\theoremstyle{remark}

\title{Vertex-Edge Weighted Molecular Graphs: A study on topological indices and their relevance to physicochemical properties of drugs in use cancer treatment   }

\author{Sezer Sorgun$^{a,*}$, Kahraman Birgin$^{a}$  \\
	\small $^{a}$Dept. of Mathematics, Nevşehir Hacı Bektaş Veli University, Nevşehir, Turkey \\
	\small $^{*}$Corresponding author \tt{ssorgun@nevsehir.edu.tr;srgnrzs@gmail.com} \\
	\small \tt{kahramanbirgin@gmail.com} \\
}

\begin{document}
	
	\maketitle
	
	\begin{abstract}
		\noindent
		
		Quantitative Structure-Property Relationship (QSPR) analysis plays a crucial role in predicting physicochemical properties and biological activities of pharmaceutical compounds, aiding in drug design and optimization. This study focuses on leveraging QSPR within the framework of vertex and edge weighted (VEW) molecular graphs, exploring their significance in drug research. By examining 48 drugs in used in the treatment of various cancers and their physicochemical properties, previous studies serve as a foundation for our research. Introducing a novel methodology for computing vertex and edge weights, exemplified by the drug Busulfan, we highlight the importance of considering atomic properties and inter-bond dynamics. Statistical analysis, employing linear regression models, reveals enhanced correlations between topological indices and physicochemical properties of drugs. Comparison with previous studies on unweighted molecular graphs highlights the enhancements achieved with our approach.

	\end{abstract}
	
	\section{Introduction}

    In chemical terminology, a molecular or chemical graph is a labeled graph in which vertices represent atoms, and edges represent the bonds between any two atoms in the molecule. Usually, hydrogen atoms are not considered in the molecular graph \cite{book}. \\
	
	A topological index (also known as a theoretical molecular descriptor) is a numerical value that plays an important role in chemical graphs, helping to estimate various physical properties and biological activities of chemical structures. Some well-known degree-based topological indices of a graph $G$ are provided in Table \ref{A}.
	
	\begin{table}[H] 
		\caption{Some degree based topological indices of a graph $G$ }
		\label{A}
		\centering
		\resizebox{\textwidth}{!}
		{\begin{tabular}{lll}
				\hline
				Topological Index (TI)  &Formula &References \\ \hline \hline
				Zagreb Index ($M_1(G)$) &$\sum_{ij\in E} (d_i+d_j)$ & Gutman and Das \cite{Gutman1} \\ \hline
				Second Zagreb Index ($M_2(G)$) &$\sum_{ij\in E} (d_id_j)$&  Das and Gutman \cite{Gutman2}  \\ \hline
				Hyper Zagreb  index ($HM(G)$)&$\sum_{ij\in E} (d_i+d_j)^2$ & Shirdel et. al. \cite{hyper} \\ \hline
				Atom bond connectivity index ($ABC(G)$)&$\sum_{ij\in E}\sqrt{\frac{d_i+d_j+2}{d_id_j}}$ & Estrada et. al. \cite{Estrada}\\ \hline
				Geometric Aritmetic index ($GA(G)$)&$\sum_{ij\in E}\frac{2\sqrt{d_id_j}}{d_i+d_j}$  & Vukicevi\'c and Furtula \cite{GA} \\ \hline
				Harmonic  index ($H(G)$)&$\sum_{ij\in E}\frac{2}{d_i+d_j}$& Fajtlowicz \cite{Harmonic}\\ \hline
				Randi\'{c}  index ($RA(G)$)&$\sum_{ij\in E}\sqrt{\frac{1}{d_id_j}}$ & Randi\'c \cite{Randic1975}\\ \hline
				Forgotten  index ($F(G)$)&$\sum_{ij\in E}(d_{i}^2+d_{j}^2)$ &  Furtula and Gutman \cite{for}\\ \hline
				Sum connectivity  index ($SC(G)$)& $\sum_{ij\in E}\sqrt{\frac{1}{d_i+d_j}}$& Gutman \cite{gutman}\\ \hline

		\end{tabular}}
	\end{table}
	
	In recent years, topological indices of graphs have become a focal point in Quantitative Structure-Property Relationship (QSPR) and Quantitative Structure Activity Relationship (QSAR) studies. These indices provide numerical representations of the structural features of chemical compounds, particularly in the context of molecular graphs (see also \cite{z1,z2}
	
	Researchers use these indices to establish mathematical relationships between molecular structure and corresponding physicochemical properties or activities, aiding in the design and optimization of new compounds for specific purposes. They continue to explore and develop new topological indices, refining their application in the field of computational chemistry and drug discovery. The ultimate goal for drugs is to utilize this knowledge to design and develop new drugs with improved efficacy and fewer side effects. The correlation between molecular structure and drug properties allows researchers to make informed decisions during the drug development process, leading to more efficient and targeted drug discovery. Typically, linear regression models are employed to establish correlations. (See also, \cite{reg1,reg2,reg3})
	
	The QSPR analysis establishes a robust correlation with topological indices, particularly concerning the physicochemical properties crucial for treating some diseases. In this analysis, structural components in the drugs represent vertices, while the bonds connecting atoms are referred as edges. Regression analysis has been employed for precise calculations in the studies. Software packages such as Statistix, Python, and MATLAB are employed to derive the results in these studies.(See also \cite{z3,z4}

	The relationships between certain physicochemical properties of drugs and their corresponding topological indices have been intensively studied using basic regression models. Efforts to combat cancer have persisted for decades, leading to significant advancements in understanding the disease and developing treatments, with ongoing work in the pharmaceutical industry. QSPR analyses of cancer drugs within the field of chemical graph theory have garnered great interest. Nasir et al. \cite{Nayir} establish the relationship between degree-based topological indices of novel drugs used in blood cancer treatment and their physicochemical properties using a linear regression model. Similarly, Shanmukha et al. \cite{Shanmukha} obtain correlations between topological indices and physicochemical properties of anticancer drugs. Bokhary et al. \cite{can} also provide a QSPR analysis for breast cancer treatment. The studies by Huang et al. \cite{can1} and Zaman et al. \cite{zaman} focus on QSPR analysis of drugs for cancer treatment. In these five studies, a basic linear regression model is employed for drug analysis as follows.
	\begin{eqnarray}\label{reg}
		P=a+b(TI)
	\end{eqnarray}
	Here, $P$ represents the physicochemical property. The variables $a$, $TI$ and $b$ denote the topological index, constant, and regression coefficient, respectively. Furthermore, the chemical graphs of the drugs given in these five studies are unweighted graphs. 
	
	Drawing inspiration from these studies, we may pose the following motivational research question: \\
	\textbf{Question 1.} Might enhanced correlations be achieved through QSPR analysis tailored to the atomic properties and inter-bond dynamics within these pharmaceutical compounds?"
	
	For the answer to this question, we need a deeper understanding of the vertex-edge weighted graph.
	
	A vertex and edge weighted (VEW) molecular graph is defined in \cite{book,hand} as  $G =G(V,E,Sym,Bo,Vw,Ew,w)$ such that the vertex and edge set is $V= V(G)$ and $E =E(G)$, respectively; a set of chemical symbols of the vertices $Sym=Sym(G)$, a set of topological bond orders ( takes the value 1 for single bonds, 2 for double bonds, 3 for triple bonds and 1.5 for aromatic bonds) of the edges $Bo = Bo(G)$, a vertex weight set $Vw(w) = V_w(w,G)$, and an edge weight set $Ew(w) = E_w(G)$. Here $w$ is the weighting scheme which is used to compute the $Vw(w)$ and $Ew(w)$. Generally, all schemes  in a molecular graph are the  properties of the atoms such as atomic number, atomic radius etc. \cite{hand}. Also, the vertex and edge weights depend on the carbon atom in the molecules and are defined as follows:
	
	\begin{equation} \label{vweighted}
		Vw(w)_{i}=1-\frac{w_{C}}{w_{i}}
	\end{equation}
	and 
	\begin{equation} \label{edgeweighted}
		Ew(w)_{ij}=\frac{w_{C}w_{C}}{Bo_{ij}w_{i}w_{j}}
	\end{equation}
	where $Vw(w)_{i}$ represents atom $i$ from a molecule; $Ew(w)_{ij}$ represents the bonds between atom $i$ and atom $j$ and $Bo_{ij}$ is the topological bonds order between $i$ and $j$, respectively \cite{book}.  
	Barysz et.al. \cite{Tri} applied a method which  based on the atomic number and the topological bond order for the  computation of topological indices.
	
	\begin{table}[H] 
		\caption{The atomic properties of some elements (Values from both PubChem and SpyderChem databases have been utilized.)}
		\label{atom}
		\begin{center}
			\resizebox{\textwidth}{!}
			{
				\begin{tabular}{|l||c|c|c|c|c|c|c|c|c|}
					\hline
					Property & $C$ & $N$ & $O$ & $F$ & $S$& $Cl$ & $Pt$  & $Br$ & $P$ \\ \hline \hline
					Atomic Number & $6$ & $7$ & $8$ & $9$ &$16$ & $17$ & $78$ & $35$  & $15$\\
					\hline  
					Atomic Mass & $12.011$ & $14.007$ & $15.999$ & $18.998$ & $32.066$ & $35.453$& $195.079$ & $79.904$ & $30.974$ \\
					\hline  
					Atomic Radius (pm) & $70$ & $65$ & $60$ & $50$ & $100$ & $100$& $135$ & $115$ & $100$\\
					\hline  
					Electro-negativity & $2.55$ & $3.12$ & $3.62$ & $4.23$ & $2.49$ & $2.82$& $2.28$ & $2.56$ & $2.22$ \\
					\hline  
					Atomic Radius (Van Der Waals) & $170$ & $155$ & $152$ & $135$ & $180$ & $175$& $209$ & $183$  & $180$\\
					\hline
					
					\hline \hline
					
			\end{tabular}}
		\end{center}
		\label{atomproperties}
	\end{table}

	 In addition, the chemical properties of drugs used in the studies are presented in a combined form in Table \ref{B}. The values in Table \ref{B} are derived from the references \cite{Nayir,Shanmukha,can,can1,zaman} as well as the PubChem database. Table \ref{B} illustrates the physicochemical properties of 48 drugs used in cancer treatment.

	\begin{table}[htbp]
		\centering
		\caption{Chemical properties of some cancer drugs}
		\label{B}
		\resizebox{\textwidth}{!}{
			\begin{tabular}{llcccccc}
				\toprule
				No.&Drug Name& Boiling Point (BP)    & Melting Point (MP)    & Entalphy (E)         & Molar Refractivity (MR)    & Molar Volume (MV)    & Complexity (C) \\
					& & $^\circ C$    &  $^\circ C$    & $kj/mol$        & $cm^3$   & $cm^3$    &  \\
				\midrule
				1&Azacitidine & 534.21 & 229.00 & 93.2  & 51.1 & 117.1 & 384.0 \\
				2&Busulfan & 464.00 & 118.00 & 69.8  & 50.9 & 182.4 & 294.0 \\
				3&Mercaptopurine & 491.00 & 313.00 & 72.8  & 41.0 & 94.2 & 190.0 \\
				4&Tioguanine & 460.70 & NaN & NaN  & NaN & 80.2 & 225.0 \\
				5&Nelarabine & 721.00 & NaN & 110.6  & 65.8 & 149.9 & 377.0 \\
				6&Cytarabine & 547.70 & 212.50 & 94.8  & 52.0 & 128.4 & 383.0 \\
				7&Clofarabine & 550.00 & NaN & 93.9  & 63.6 & 143.1 & 370.0 \\
				8&Bosutinib & 649.70 & NaN & 95.8  & 141.9 & 388.3 & 734.0 \\
				9&Dasatinib & 133.08 & NaN & NaN  & 132.0 & 366.4 & 642.0 \\
				10&MelphalaN & 473.00 & NaN & 77.6  & 78.8 & 231.2 & 265.0 \\
				11&Dexamethasone & 568.20 & 263.00 & 98.0 & 100.2 & 296.2 & 805.0 \\
				12&Doxorubicine & 216.00 & 204.50 & 123.5  & 131.5 & 336.6 & 977.0 \\
				13&Carboplatin & 60.04 & 200.00 & NaN & NaN & 366.4 & 177.0 \\
				14&Amathaspiramide & 572.70 & 209.72 & 90.3  & 89.4 & 233.9 & NaN \\
				15&Aminopterin & 782.27 & 344.45 & NaN  & 114.0 & 277.2 & 674.0 \\
				16&Aspidostomide & 798.80 & NaN & 116.2 & 116.0 & 262.0 & NaN \\
				17&Carmustine & 309.60 & 120.99 & 63.8 & 46.6 & 146.4 & 156.0 \\
				18&Caulibugulone & 373.00 & 129.46 & 62.0  & 52.2 & 139.1 & 351.0 \\
				19&ConvolutamideA & 629.90 & NaN & 97.9 & 130.1 & 396.0 & NaN \\
				20&ConvolutamineF & 387.70 & 128.67 & 63.7  & 73.8 & 220.1 & 194.0 \\
				21&Convolutamydine & 504.90 & 199.20 & 81.6  & 68.2 & 190.0 & NaN \\
				22&Daunorubicin & 770.00 & 208.50 & 117.6 & 130.0 & 339.4 & NaN \\
				23&Deguelin & 560.10 & 213.39 & 84.3  & 105.1 & 314.2 & 674.0 \\
				24&Melatonin & 512.80 & 118.00 & 78.4  & 67.6 & 197.6 & 270.0 \\
				25&Minocycline & 803.30 & 326.30 & 122.5  & 116.0 & 294.6 & 971.0 \\
				26&Perfragilin & 431.50 & 187.62 & 68.7 & 63.6 & 167.8 & 558.0 \\
				27& Podophyllotoxin & 597.90 & 235.86 & 93.6  & 104.3 & 302.4 & 628.0 \\
				28&Pterocellin & 521.60 & 199.88 & 79.5  & 87.4 & 228.3 & 618.0 \\
				29&Raloxifene & 728.20 & 289.58 & 110.1  & 136.6 & 367.3 & 655.0 \\
				30&Tambjamine & 391.70 & NaN & 64.1 & 76.6 & 235.1 & 410.0 \\
				31&Abemaciclib & 689.30 & NaN & 101.0 & 140.4 & 382.3 & 723.0 \\
				32&Abraxane & 957.10 & NaN & 146.0  & 219.3 & 610.6 & 1790.0 \\
				33&Anastrozole & 469.70 & 81.50 & 73.2 & 90.0 & 270.3 & 456.0 \\
				34&Capecitabine & 517.00 & 115.50 & NaN  & 82.3 & 240.5 & 582.0 \\
				35&Cyclophosphamide & 336.10 & 51.00 & 57.9  & 58.1 & 195.7 & 212.0 \\
				36&Everolimus & 998.70 & 998.70 & 165.1  & 257.7 & 811.2 & 1810.0 \\
				37&Exemestane & 453.70 & 155.13 & 71.3  & 85.8 & 260.6 & 653.0 \\
				38&Fulvestrant & 674.80 & 104.00 & 104.1  & 154.0 & 505.1 & 854.0 \\
				39&Ixabepilone & 697.80 & NaN & 107.3 & 140.1 & 451.6 & 817.0 \\
				40&Letrozole & 563.50 & 181.00 & 84.7  & 87.1 & 234.5 & 420.0 \\
				41&Megestrol Acetate & 507.10 & 214.00 & 77.7  & 106.4 & 333.4 & 821.0 \\
				42&Methotrexate & NaN & 192.00 & NaN  & 119.0 & 295.7 & 704.0 \\
				43&Tamoxifen & 482.30 & 96.00 & 74.7  & 118.9 & 118.9 & 463.0 \\
				44&Thiotepa & 270.20 & 51.50 & 50.8  & 49.1 & 125.8 & 194.0 \\
				45&Glasdegib & 633.40 & NaN & 93.6  & 106.9 & 281.0 & 595.0 \\
				46&Palbociclib & 711.50 & NaN & 104.0  & 123.9 & 340.7 & 775.0 \\
				47&Gilteritinib & 696.90 & NaN & 102.1  & 157.8 & 444.9 & 785.0 \\
				48&Ivosidenib & 854.30 & NaN & 124.1  & 140.1 & 383.6 & 1050.0 \\
				\bottomrule
		\end{tabular}}
	\end{table}

	We now possess the information necessary to address Question 1. Therefore, the second section of this paper will present a method for calculating the vertex and edge weights for the drugs listed in Table \ref{B}. For this purpose, the drug Busulfan was selected as a model to provide a more detailed understanding, and vertex-edge weights were calculated based on the properties of the atoms, particularly the atomic radius. Section 3 contains a statistical analysis presenting a linear regression model with Molar Refractivity ($MR$) as the dependent variable and the topological indices of drugs as independent variables for the atomic radius scheme. Additionally, we present all correlations between the physicochemical properties of the drugs and the topological indices within the same scheme. In Section 4, we compare the QSPR analyses conducted in previous studies using unweighted molecular graphs of drugs with the analyses performed using our new approach.


	\section{Material and Method }
	
In edge-weighted graphs, the degree of a vertex is determined as the sum of the weights of all edges incident to the vertex. It's noteworthy that, in cases where there are no vertex-weights, they are considered as zero. However, in vertex-edge weighted graphs, considering the weight of the vertex becomes pertinent in defining the vertex's degree. Hence, the degree of a vertex (atom) $i$ in the $VEW$ molecular graph can be calculated using equations (\ref{vweighted}) and (\ref{edgeweighted}), as follows: 
	\begin{eqnarray} \label{degree}
		\Lambda_i= Vw(w)_{i}+\sum_{ij\in E} Ew(w)_{ij}
	\end{eqnarray}
By substituting the values of $\Lambda_i$ into equation (\ref{degree}) in place of the $d_i$'s in Table \ref{A}, we can systematically organize the topological indices of the $VEW$ molecular graphs. \\
	
	For further clarification, let's calculate the vertex and edge weights for a drug and determine the degree of each vertex. Consider the drug Busulfan in Table \ref{B}. Its chemical structure depiction and its labeled graph with atomic radius are shown in Figure 1. Recalling the equation in (1), the vertex weights of each atoms for the scheme of atomic radius in Table \ref{atom} are the following
	\begin{eqnarray*}
	Vw(w)_1&=&Vw(w)_C=1-\frac{w_C}{w_C}=1-\frac{70}{70}=0\\ \nonumber
	Vw(w)_2&=&Vw(w)_S=1-\frac{w_C}{w_S}=1-\frac{70}{100}=0.3\\ \nonumber
	Vw(w)_5&=&Vw(w)_O=1-\frac{w_C}{w_O}=1-\frac{70}{60}=-0.166
	\end{eqnarray*}
	
	\begin{figure}[htbp]
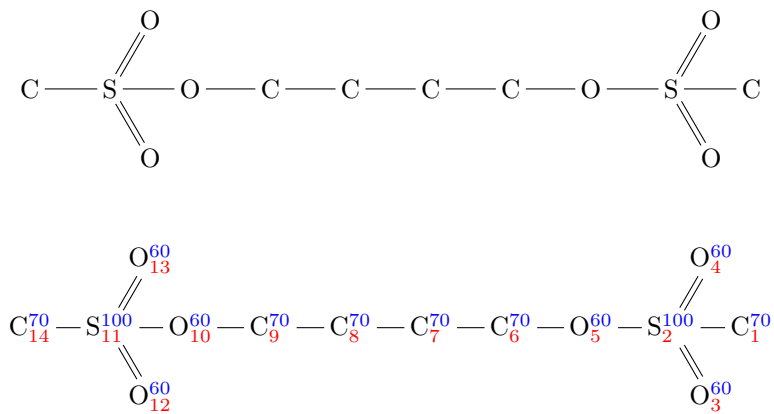

		\centering
		\chemfig{S (=[::+60]O)(=[::-60]O)(-[:180]C)-O-C-C-C-C-O-S(=[::+60]O)(=[::-60]O)-C} \\ \vspace{1cm}
		\chemfig{S_{\textcolor{red}{11}}^{\textcolor{blue}{100}}(=[::+60]O_{\textcolor{red}{13}}^{\textcolor{blue}{60}})(=[::-60]O_{\textcolor{red}{12}}^{\textcolor{blue}{60}})(-[:180]C_{\textcolor{red}{14}}^{\textcolor{blue}{70}})-O_{\textcolor{red}{10}}^{\textcolor{blue}{60}}-C_{\textcolor{red}{9}}^{\textcolor{blue}{70}}-C_{\textcolor{red}{8}}^{\textcolor{blue}{70}}-C_{\textcolor{red}{7}}^{\textcolor{blue}{70}}-C_{\textcolor{red}{6}}^{\textcolor{blue}{70}}-O_{\textcolor{red}{5}}^{\textcolor{blue}{60}}-S_{\textcolor{red}{2}}^{\textcolor{blue}{100}}(=[::+60]O_{\textcolor{red}{4}}^{\textcolor{blue}{60}})(=[::-60]O_{\textcolor{red}{3}}^{\textcolor{blue}{60}})-C_{\textcolor{red}{1}}^{\textcolor{blue}{70}}}
		\caption{Busulfan with a 2D depiction and labeled atoms according to their atomic radii.}
	\end{figure}
	
	So, from the labeled graph in Figure 1, we obtain the vertex weights as follows
	
	\begin{center}
		\begin{tabular}{ccccccccccccccc}
			\hline
			$Vw(w)_1$ & $Vw(w)_2$& $Vw(w)_3$ & $Vw(w)_4$ & $Vw(w)_5$ & $Vw(w)_6$&  $Vw(w)_7$ \\
			\hline
			0 &  0.3& -0.166  & -0.166& -0.166 & 0 & 0 \\
			\hline
		\end{tabular} \\ \vspace{0.3cm}
		\begin{tabular}{ccccccccc}
			\hline
			$Vw(w)_8$ & $Vw(w)_9$ & $Vw(w)_{10}$ & $Vw(w)_{11}$ & $Vw(w)_{12}$ &$Vw(w)_{13}$ &$Vw(w)_{14}$\\
			\hline
			0 & 0& -0.166&  0.3 & -0.166 & -0.166  & 0
		\end{tabular} 
	\end{center}
	and considering the number of interatomic bonds, we obtain the following values for edge weights rounded to three decimals using equation (2). 
	\begin{center}
		\begin{tabular}{ccccccc}
			\hline
			$Ew(w)_{1,2}$& $Ew(w)_{2,3}$& $Ew(w)_{2,4}$ & $Ew(w)_{2,5}$ & $Ew(w)_{5,6}$& $Ew(w)_{6,7}$& $Ew(w)_{7,8}$ \\
			\hline
			0.7 & 0.408 & 0.408 & 0.816 & 1.166 & 1 & 1 \\
			\hline
		\end{tabular}\vspace{0.3cm}

		\begin{tabular}{cccccc}
			\hline
			$Ew(w)_{8,9}$ & $Ew(w)_{9,10}$ &$Ew(w)_{10,11}$&$Ew(w)_{11,12}$& $Ew(w)_{11,13}$&$Ew(w)_{11,14}$\\
			\hline
			1 & 1.166 & 0.816 & 0.408 & 0.408 &  0.7\\
			\hline
		\end{tabular} 
	\end{center}
	We are now prepared to calculate the degrees of all vertices. By utilizing Equation (\ref{degree}) and referring to Figure 1, we determine the degree of each atom, rounded to three decimals:
	
	\begin{center}
		\begin{tabular}{cccccccccc}
			\hline
			$\Lambda_1$ & $\Lambda_2$& $\Lambda_3$& $\Lambda_4$& $\Lambda_5$& $\Lambda_6$& $\Lambda_7$& $\Lambda_8$& $\Lambda_9$& $\Lambda_{10}$\\
			\hline
			0.7 & 2.632 & 0.242& 0.242& 1.816& 2.166 & 2 & 2  & 2.166  &1.816\\
			\hline
		\end{tabular}\vspace{0.5cm}
		
		\begin{tabular}{cccccc}
			\hline
			$\Lambda_{11}$ & $\Lambda_{12}$& $\Lambda_{13}$& $\Lambda_{14}$ \\
			\hline
			2.632 & 0.242 & 0.242& 0.7 \\
			\hline
		\end{tabular}
	\end{center}
	Then, we can easily compute the degree-based topological indices of the $VEW$ molecular graph for all 48 drugs, using any scheme provided in Table \ref{atom}.

	\section{Results and Discussion}
	
	Recalling the linear regression model in Equation (\ref{reg}), we designate the dependent variable as Molar Refractivity ($MR$) and the independent variables as the topological indices. Using Python, and applying the linear regression model for the atomic radius scheme, we obtain the following results:\\

	\begin{center}
		\begin{verbatim}
			OLS Regression Results                            
			=====================================================================
			Dep. Variable:                MR   R-squared:                   0.964
			Model:                       OLS   Adj. R-squared:              0.963
			Method:            Least Squares   F-statistic:                 1172.
			Date:           Sun, 03 Dec 2023   Prob (F-statistic):       2.37e-33
			Time:                   17:49:02   Log-Likelihood:            -162.63
			No. Observations:             46   AIC:                         329.3
			Df Residuals:                 44   BIC:                         332.9
			Df Model:                           1                                         
			Covariance Type:            nonrobust                                         
			=====================================================================      


			coef    std err     t        P>|t|      [0.025      0.975]
			------------------------------------------------------------------
			const     3.7481   3.149     1.190      0.24       -2.599      10.095
			SC Index  7.2615   0.212    34.232      0.000       6.834       7.689
			==================================================================
			Omnibus:                   2.178   Durbin-Watson:              1.702
			Prob(Omnibus):             0.337   Jarque-Bera (JB):           1.587
			Skew:                      0.253   Prob(JB):                   0.452
			Kurtosis:                  2.244   Cond. No.                   37.5
			=====================================================================
			Correlation with Dependent Variable (MR):
			const            NaN
			SC Index    0.981738
		\end{verbatim}
	\end{center}
	
	From the regression results, it is evident that approximately $96.5$ percent of the variance in $MR$ is explained by $SC$ index, suggesting a very strong relationship. Additionally, since the p-value for the $SC$ index is very low, the result indicates a highly significant relationship. In summary, the results of the analysis appear to indicate a robust model with a high R-squared value and a significant relationship between the $SC$ index and $MR$, with respect to the scheme of atomic radius.
		
	Similarly, we present the correlations with other dependent variables for all drugs in Table \ref{S2}. Additionally, Table \ref{S1} provides all degree-based topological indices of the drugs with respect to the atomic radius of the atoms. 
		
	\begin{center}
	\begin{table}[htbp]
	\caption{The topological indices (TIs) of the drugs with respect to the scheme of atomic radius.}
	\label{S1}
	\resizebox{\textwidth}{!}
	{\begin{tabular}{|l|ccccccccc|}
			\hline
			\textbf{Drug Name} & \textbf{$M_1$ Index} & \textbf{$M_2$ Index} & \textbf{ABC Index} & \textbf{Harmonic Index} & \textbf{Hyperzagreb Index} & \textbf{$RA$ Index} & \textbf{$GA$ Index} & \textbf{Forgotten Index} & \textbf{$SC$ Index} \\
			\hline
			Azacitidine & 91.29 & 113.18 & 20.93 & 7.43 & 481.29 & 8.08 & 17.03 & 254.93 & 8.13 \\
			Busulfan & 47.37 & 36.34 & 22.87 & 7.35 & 177.35 & 9.87 & 10.81 & 104.67 & 6.88 \\
			Mercaptopurine & 51.85 & 61.17 & 12.73 & 4.82 & 251.75 & 5.00 & 10.74 & 129.40 & 5.13 \\
			Tioguanine & 58.24 & 68.88 & 13.85 & 5.10 & 290.19 & 5.38 & 11.55 & 152.44 & 5.51 \\
			Nelarabine & 120.74 & 156.59 & 25.14 & 9.11 & 654.64 & 9.52 & 22.23 & 341.46 & 10.18 \\
			Cytarabine & 90.76 & 111.98 & 21.01 & 7.48 & 476.14 & 8.13 & 17.03 & 252.18 & 8.15 \\
			Clofarabine & 115.54 & 149.51 & 24.13 & 8.69 & 626.72 & 9.13 & 21.19 & 327.70 & 9.73 \\
			Bosutinib & 188.29 & 225.86 & 45.53 & 17.09 & 940.82 & 17.95 & 37.83 & 489.11 & 18.09 \\
			Dasatinib & 172.12 & 200.58 & 41.59 & 15.44 & 841.40 & 16.19 & 34.81 & 440.25 & 16.62 \\
			MelphalaN & 84.70 & 93.16 & 23.75 & 9.01 & 395.64 & 9.59 & 18.17 & 209.32 & 9.18 \\
			Dexamethasone & 162.93 & 209.40 & 36.78 & 12.82 & 925.79 & 14.22 & 28.72 & 506.99 & 13.95 \\
			Doxorubicine & 220.90 & 276.15 & 49.92 & 17.61 & 1180.48 & 19.24 & 40.61 & 628.18 & 19.32 \\
			Carboplatin & 47.33 & 53.79 & 13.37 & 4.59 & 243.31 & 5.38 & 8.97 & 135.72 & 4.74 \\
			Amathaspiramide & 118.78 & 145.51 & 28.27 & 10.33 & 621.60 & 11.05 & 22.84 & 330.58 & 11.04 \\
			Aminopterin & 161.50 & 184.66 & 41.09 & 14.79 & 789.07 & 16.14 & 32.15 & 419.75 & 15.79 \\
			Aspidostomide & 143.68 & 178.08 & 33.37 & 12.24 & 739.70 & 12.96 & 27.95 & 383.54 & 13.24 \\
			Carmustine & 43.48 & 43.51 & 15.71 & 6.11 & 186.61 & 6.71 & 10.33 & 99.59 & 5.73 \\
			Caulibugulone & 66.13 & 72.94 & 18.76 & 7.16 & 304.60 & 7.60 & 14.43 & 158.72 & 7.28 \\
			ConvolutamideA & 146.23 & 164.11 & 39.37 & 14.68 & 701.85 & 15.71 & 30.53 & 373.64 & 15.24 \\
			ConvolutamineF & 67.86 & 76.53 & 18.05 & 6.92 & 319.55 & 7.20 & 14.52 & 166.50 & 7.16 \\
			Convolutamydine & 87.27 & 102.02 & 22.41 & 7.89 & 447.94 & 8.86 & 16.67 & 243.90 & 8.36 \\
			Daunorubicin & 216.56 & 270.97 & 48.75 & 17.11 & 1160.73 & 18.76 & 39.58 & 618.79 & 18.83 \\
			Deguelin & 170.17 & 212.58 & 36.95 & 13.34 & 905.35 & 14.10 & 31.68 & 480.19 & 14.75 \\
			Melatonin & 84.24 & 97.14 & 21.37 & 7.96 & 406.42 & 8.43 & 17.35 & 212.15 & 8.43 \\
			Minocycline & 180.59 & 219.76 & 43.03 & 15.10 & 947.12 & 16.74 & 33.66 & 507.61 & 16.38 \\
			Perfragilin & 80.85 & 89.17 & 23.88 & 8.58 & 383.46 & 9.73 & 16.65 & 205.13 & 8.71 \\
			Podophyllotoxin & 175.17 & 225.13 & 37.45 & 13.69 & 929.32 & 14.28 & 33.10 & 479.06 & 15.18 \\
			Pterocellin & 130.95 & 158.99 & 31.50 & 11.61 & 658.13 & 12.33 & 26.11 & 340.15 & 12.45 \\
			Raloxifene & 180.86 & 210.80 & 43.57 & 16.32 & 878.07 & 16.95 & 36.99 & 456.46 & 17.56 \\
			Tambjamine & 92.46 & 105.33 & 23.44 & 8.96 & 441.68 & 9.24 & 19.45 & 231.02 & 9.42 \\
			Abemaciclib & 208.32 & 257.06 & 45.22 & 16.53 & 1081.77 & 17.19 & 39.68 & 567.64 & 18.35 \\
			Abraxane & 340.95 & 418.78 & 80.84 & 28.66 & 1801.85 & 31.46 & 64.04 & 964.30 & 30.99 \\
			Anastrozole & 111.95 & 128.15 & 29.69 & 10.68 & 577.42 & 12.07 & 21.23 & 321.11 & 10.85 \\
			Capecitabine & 127.34 & 152.44 & 31.15 & 11.14 & 650.90 & 12.18 & 24.52 & 346.02 & 11.96 \\
			Cyclophosphamide & 57.60 & 58.43 & 18.84 & 7.13 & 247.33 & 7.79 & 13.34 & 130.47 & 7.02 \\
			Everolimus & 331.41 & 378.85 & 87.40 & 32.16 & 1626.81 & 34.80 & 67.22 & 869.11 & 33.55 \\
			Exemestane & 121.83 & 143.90 & 30.80 & 10.97 & 630.44 & 12.17 & 23.31 & 342.64 & 11.61 \\
			Fulvestrant & 222.37 & 267.24 & 51.79 & 18.52 & 1197.92 & 20.06 & 41.43 & 663.44 & 20.03 \\
			Ixabepilone & 181.46 & 210.24 & 44.52 & 15.80 & 930.17 & 17.35 & 34.49 & 509.69 & 16.99 \\
			Letrozole & 110.03 & 128.90 & 30.57 & 11.68 & 533.69 & 12.63 & 23.16 & 275.89 & 11.62 \\
			Megestrol Acetate & 155.00 & 189.17 & 37.74 & 13.30 & 828.93 & 14.79 & 28.83 & 450.58 & 14.23 \\
			Methotrexate & 167.88 & 193.61 & 42.16 & 15.10 & 829.34 & 16.52 & 33.02 & 442.12 & 16.18 \\
			Tamoxifen & 136.94 & 153.82 & 34.77 & 13.41 & 636.78 & 13.68 & 29.45 & 329.15 & 14.15 \\
			Thiotepa & 62.17 & 73.58 & 14.90 & 5.55 & 302.85 & 5.79 & 12.67 & 155.69 & 5.99 \\
			Glasdegib & 147.55 & 173.69 & 37.16 & 13.83 & 726.78 & 14.78 & 29.87 & 379.41 & 14.49 \\
			Palbociclib & 181.26 & 220.86 & 42.54 & 15.66 & 916.70 & 16.53 & 35.82 & 474.98 & 16.94 \\
			Gilteritinib & 215.71 & 260.92 & 49.72 & 18.51 & 1085.29 & 19.22 & 42.82 & 563.45 & 20.09 \\
			Ivosidenib & 221.92 & 263.68 & 54.37 & 19.52 & 1144.96 & 21.42 & 42.45 & 617.59 & 20.74 \\
			\hline
	\end{tabular}}
\end{table}
\begin{table}[htbp]
	\caption{Correlations for the scheme of atomic radius}
	\label{S2}
	\resizebox{\textwidth}{!}
	{	\begin{tabular}{lccccccccc}
			\hline
			Variable & $MR$ & $BP$ & $MP$ & $Enthalpy$ & $MV$ & $Complexity$ \\
			\hline
			M1 Index & 0.954707 & 0.692002 & 0.646851 & 0.877486 & 0.859001 & 0.950021 \\
			M2 Index & 0.927417 & 0.686650 & 0.612071 & 0.876430 & 0.830374 & 0.938073 \\
			ABC Index & 0.976444 & 0.689719 & 0.708682 & 0.863176 & 0.895718 & 0.965452 \\
			Harmonic Index & 0.984422 & 0.687472 & 0.717517 & 0.853794 & 0.895916 & 0.954488 \\
			Hyperzagreb Index & 0.927055 & 0.684624 & 0.606496 & 0.875562 & 0.838934 & 0.944608 \\
			RA Index & 0.975424 & 0.685852 & 0.717832 & 0.854366 & 0.899163 & 0.963791 \\
			GA Index & 0.975732 & 0.692292 & 0.682079 & 0.867154 & 0.872540 & 0.944493 \\
			Forgotten Index & 0.924794 & 0.681484 & 0.600446 & 0.872990 & 0.844698 & 0.948386 \\
			SC Index & 0.981738 & 0.691492 & 0.702606 & 0.864488 & 0.888248 & 0.955267 \\
			\hline
	\end{tabular}
}
   \end{table}
\end{center}

It is possible to make further inferences from Table \ref{S2}. For example, MR and Complexity have a strong correlation with the topological indices while $Entalphy$ and $MV$ have an average correlation. Similarly, we compute all the topological indices  in terms of all other schemes and again by the same regression, we have the Table \ref{S3}-Table \ref{S12}.

\begin{center}
\begin{table}[htbp]
	\caption{The topological indices (TIs) of the drugs with respect to the scheme of of atomic mass}
	\label{S3}
	\resizebox{\textwidth}{!}
	{\begin{tabular}{|l|ccccccccc|}
			\hline
			\textbf{Drug Name} & \textbf{$M_1$ Index} & \textbf{$M_2$ Index} & \textbf{ABC Index} & \textbf{Harmonic Index} & \textbf{Hyperzagreb Index} & \textbf{RA Index} & \textbf{GA Index} & \textbf{Forgotten Index} & \textbf{SC Index} \\
			\hline
			Azacitidine & 77.20 & 81.67 & 22.59 & 8.75 & 343.27 & 9.18 & 17.34 & 179.92 & 8.83 \\
			Busulfan & 34.94 & 23.10 & 25.27 & 10.46 & 101.44 & 12.06 & 11.74 & 55.24 & 8.17 \\
			Mercaptopurine & 45.38 & 47.20 & 13.73 & 5.52 & 193.58 & 5.63 & 10.83 & 99.18 & 5.49 \\
			Tioguanine & 50.67 & 52.96 & 14.88 & 5.86 & 220.06 & 6.05 & 11.70 & 114.15 & 5.91 \\
			Nelarabine & 102.82 & 114.47 & 27.80 & 10.75 & 475.97 & 11.15 & 22.35 & 247.03 & 11.05 \\
			Cytarabine & 78.20 & 83.65 & 22.40 & 8.63 & 351.59 & 9.07 & 17.33 & 184.30 & 8.77 \\
			Clofarabine & 97.97 & 108.61 & 26.50 & 10.26 & 451.14 & 10.61 & 21.41 & 233.93 & 10.57 \\
			Bosutinib & 169.88 & 185.52 & 48.44 & 19.01 & 770.76 & 19.74 & 37.93 & 399.71 & 19.08 \\
			Dasatinib & 153.15 & 160.46 & 44.22 & 17.36 & 666.58 & 17.88 & 35.13 & 345.65 & 17.62 \\
			MelphalaN & 78.08 & 81.03 & 24.69 & 9.92 & 339.08 & 10.28 & 18.45 & 177.03 & 9.61 \\
			Dexamethasone & 155.29 & 191.22 & 36.88 & 13.38 & 835.20 & 14.42 & 29.15 & 452.76 & 14.26 \\
			Doxorubicine & 203.60 & 238.57 & 51.39 & 19.17 & 1006.91 & 20.33 & 41.19 & 529.77 & 20.15 \\
			Carboplatin & 44.01 & 47.92 & 13.29 & 5.05 & 213.74 & 5.50 & 9.33 & 117.90 & 4.96 \\
			Amathaspiramide & 107.70 & 121.51 & 29.87 & 11.50 & 515.45 & 12.10 & 23.05 & 272.43 & 11.64 \\
			Aminopterin & 146.35 & 154.86 & 42.36 & 16.38 & 649.24 & 17.17 & 32.82 & 339.52 & 16.61 \\
			Aspidostomide & 127.09 & 142.14 & 35.74 & 14.01 & 584.94 & 14.51 & 28.27 & 300.66 & 14.15 \\
			Carmustine & 36.14 & 30.32 & 16.89 & 7.28 & 127.39 & 7.58 & 10.64 & 66.74 & 6.26 \\
			Caulibugulone & 62.97 & 66.87 & 19.16 & 7.59 & 278.45 & 7.90 & 14.54 & 144.71 & 7.48 \\
			ConvolutamideA & 136.66 & 144.41 & 39.97 & 15.58 & 607.24 & 16.24 & 30.98 & 318.42 & 15.71 \\
			ConvolutamineF & 60.14 & 60.67 & 19.66 & 7.90 & 252.89 & 8.21 & 14.55 & 131.54 & 7.64 \\
			Convolutamydine & 80.27 & 88.28 & 22.71 & 8.59 & 379.64 & 9.19 & 17.08 & 203.08 & 8.72 \\
			Daunorubicin & 200.10 & 235.04 & 50.04 & 18.55 & 993.72 & 19.73 & 40.15 & 523.64 & 19.60 \\
			Deguelin & 154.78 & 177.42 & 39.22 & 14.81 & 753.53 & 15.52 & 31.77 & 398.70 & 15.52 \\
			Melatonin & 78.26 & 85.69 & 22.28 & 8.71 & 355.22 & 9.05 & 17.51 & 183.84 & 8.79 \\
			Minocycline & 168.45 & 194.40 & 43.97 & 16.26 & 826.80 & 17.49 & 34.14 & 438.01 & 16.98 \\
			Perfragilin & 74.36 & 77.01 & 24.75 & 9.52 & 327.88 & 10.42 & 16.93 & 173.87 & 9.14 \\
			Podophyllotoxin & 157.28 & 185.99 & 40.70 & 15.75 & 766.23 & 16.31 & 33.15 & 394.26 & 16.20 \\
			Pterocellin & 123.01 & 140.61 & 32.20 & 12.36 & 579.54 & 12.84 & 26.33 & 298.33 & 12.84 \\
			Raloxifene & 169.80 & 187.59 & 44.94 & 17.40 & 775.28 & 17.85 & 37.21 & 400.10 & 18.13 \\
			Tambjamine & 85.80 & 90.67 & 24.77 & 9.72 & 380.99 & 10.05 & 19.42 & 199.65 & 9.80 \\
			Abemaciclib & 186.82 & 208.80 & 48.08 & 18.41 & 869.29 & 18.97 & 39.97 & 451.68 & 19.37 \\
			Abraxane & 315.90 & 363.55 & 82.43 & 31.03 & 1550.99 & 32.84 & 65.10 & 823.89 & 32.24 \\
			Anastrozole & 106.54 & 116.07 & 29.72 & 11.15 & 527.88 & 12.13 & 21.42 & 295.74 & 11.11 \\
			Capecitabine & 108.89 & 112.65 & 33.11 & 12.91 & 472.60 & 13.56 & 25.06 & 247.31 & 12.89 \\
			Cyclophosphamide & 45.07 & 36.67 & 21.12 & 9.07 & 150.51 & 9.41 & 13.67 & 77.18 & 7.92 \\
			Everolimus & 305.05 & 321.68 & 90.17 & 34.80 & 1369.28 & 36.78 & 67.96 & 725.93 & 34.91 \\
			Exemestane & 121.00 & 142.85 & 30.28 & 11.06 & 622.58 & 11.96 & 23.57 & 336.88 & 11.65 \\
			Fulvestrant & 201.58 & 229.47 & 53.51 & 20.41 & 973.12 & 21.43 & 42.36 & 514.19 & 21.02 \\
			Ixabepilone & 168.60 & 182.36 & 45.75 & 17.02 & 803.43 & 18.24 & 34.88 & 438.71 & 17.64 \\
			Letrozole & 104.40 & 115.66 & 30.58 & 12.14 & 479.66 & 12.68 & 23.36 & 248.33 & 11.89 \\
			Megestrol Acetate & 150.01 & 179.56 & 37.46 & 13.81 & 778.61 & 14.82 & 29.31 & 419.49 & 14.49 \\
			Methotrexate & 151.63 & 161.23 & 43.63 & 16.77 & 677.93 & 17.66 & 33.68 & 355.47 & 17.05 \\
			Tamoxifen & 131.83 & 143.59 & 35.77 & 13.97 & 592.12 & 14.29 & 29.44 & 304.94 & 14.43 \\
			Thiotepa & 49.43 & 46.99 & 16.90 & 6.97 & 190.35 & 7.08 & 12.86 & 96.38 & 6.71 \\
			Glasdegib & 137.48 & 151.57 & 37.88 & 14.72 & 629.07 & 15.31 & 30.18 & 325.93 & 14.98 \\
			Palbociclib & 166.38 & 186.88 & 44.03 & 16.99 & 770.63 & 17.54 & 36.16 & 396.87 & 17.65 \\
			Gilteritinib & 194.50 & 213.71 & 52.87 & 20.58 & 883.83 & 21.18 & 43.02 & 456.42 & 21.18 \\
			Ivosidenib & 199.96 & 219.73 & 55.50 & 21.33 & 920.54 & 22.41 & 43.44 & 481.08 & 21.73 \\
			\hline
	\end{tabular}}
\end{table}

\begin{table} [htbp]
\caption{Correlations for the scheme of atomic mass}
\label{S4}
\resizebox{\textwidth}{!}
{\begin{tabular}{lccccccc}
		\hline
		Variable & $MR$ & $BP$ & $MP$ & $Entalphy$ & $MV$ & $Complexity$ \\
		\hline
		$M_1$ Index & 0.956436 & 0.685722 & 0.641661 & 0.865331 & 0.860900 & 0.954453 \\
		$M_2$ Index & 0.929797 & 0.673603 & 0.593199 & 0.852423 & 0.835349 & 0.946240 \\
		ABC Index & 0.976896 & 0.691754 & 0.710028 & 0.867395 & 0.892970 & 0.957841 \\
		Harmonic Index & 0.977603 & 0.688825 & 0.716190 & 0.859685 & 0.892096 & 0.945217 \\
		Hyperzagreb Index & 0.927412 & 0.668922 & 0.593478 & 0.848835 & 0.840255 & 0.950608 \\
		RA Index & 0.973375 & 0.687963 & 0.717089 & 0.859882 & 0.894804 & 0.952140 \\
		GA Index & 0.975392 & 0.693305 & 0.681922 & 0.869379 & 0.874417 & 0.947140 \\
		Forgotten Index & 0.923947 & 0.663818 & 0.592942 & 0.844432 & 0.843438 & 0.953145 \\
		SC Index & 0.979582 & 0.693226 & 0.703425 & 0.868931 & 0.886892 & 0.951749 \\
		\hline
\end{tabular}}
\end{table}
\end{center}

\begin{center}
\begin{table}[htbp]
	\caption{The topological indices (TIs) of the drugs with respect to the scheme of electronegativity}
	\label{S5}
	\centering
	\resizebox{\textwidth}{!}
	{	\begin{tabular}{lcccccccccc}
			\hline
			{Drug} &  {SC Index} & {$M_1$ Index} & {$M_2$ Index} & {ABC Index} & {Harmonic Index} & {Hyperzagreb Index} & {RA Index} & {GA Index} & {Forgotten Index} \\
			\midrule
			Azacitidine & 75.13 & 77.49 & 22.92 & 8.99 & 325.19 & 9.40 & 17.38 & 170.20 & 8.95 \\
			Busulfan & 45.92 & 36.51 & 19.04 & 7.44 & 164.05 & 8.15 & 12.06 & 91.03 & 6.95 \\
			Mercaptopurine & 44.91 & 45.93 & 14.19 & 5.60 & 190.13 & 5.86 & 10.68 & 98.26 & 5.52 \\
			Tioguanine & 50.00 & 51.36 & 15.34 & 5.95 & 214.66 & 6.28 & 11.57 & 111.95 & 5.95 \\
			Nelarabine & 100.19 & 108.82 & 28.27 & 11.04 & 452.20 & 11.44 & 22.36 & 234.56 & 11.20 \\
			Cytarabine & 76.41 & 79.97 & 22.67 & 8.83 & 335.58 & 9.24 & 17.37 & 175.64 & 8.87 \\
			Clofarabine & 97.37 & 106.25 & 26.58 & 10.26 & 443.18 & 10.64 & 21.36 & 230.68 & 10.58 \\
			Bosutinib & 170.87 & 187.01 & 48.46 & 18.94 & 780.73 & 19.73 & 37.82 & 406.70 & 19.04 \\
			Dasatinib & 157.33 & 168.74 & 43.53 & 16.92 & 703.56 & 17.45 & 35.07 & 366.08 & 17.39 \\
			MelphalaN & 79.51 & 82.63 & 24.07 & 9.49 & 345.54 & 9.87 & 18.42 & 180.27 & 9.44 \\
			Dexamethasone & 154.66 & 189.85 & 36.91 & 13.43 & 828.54 & 14.45 & 29.19 & 448.85 & 14.29 \\
			Doxorubicine & 201.63 & 234.50 & 51.66 & 19.39 & 988.56 & 20.51 & 41.24 & 519.55 & 20.26 \\
			Carboplatin & 43.64 & 47.23 & 13.32 & 5.11 & 210.68 & 5.53 & 9.36 & 116.22 & 4.99 \\
			Amathaspiramide & 111.55 & 128.84 & 29.39 & 11.10 & 550.12 & 11.77 & 22.90 & 292.44 & 11.43 \\
			Aminopterin & 143.84 & 149.98 & 42.72 & 16.67 & 627.48 & 17.41 & 32.90 & 327.53 & 16.75 \\
			Aspidostomide & 138.21 & 163.56 & 34.13 & 12.77 & 683.80 & 13.43 & 27.94 & 356.67 & 13.51 \\
			Carmustine & 37.45 & 31.88 & 16.36 & 6.91 & 133.80 & 7.22 & 10.62 & 70.05 & 6.11 \\
			Caulibugulone & 62.42 & 65.84 & 19.28 & 7.67 & 274.17 & 7.98 & 14.55 & 142.49 & 7.52 \\
			ConvolutamideA & 140.76 & 152.44 & 39.35 & 15.14 & 644.09 & 15.83 & 30.88 & 339.20 & 15.49 \\
			ConvolutamineF & 67.28 & 75.39 & 18.75 & 7.19 & 319.69 & 7.60 & 14.34 & 168.92 & 7.27 \\
			Convolutamydine & 84.74 & 96.36 & 22.02 & 8.07 & 418.44 & 8.73 & 16.94 & 225.73 & 8.46 \\
			Daunorubicin & 198.22 & 231.14 & 50.28 & 18.75 & 976.03 & 19.90 & 40.20 & 513.74 & 19.70 \\
			Deguelin & 153.06 & 173.75 & 39.57 & 15.01 & 737.92 & 15.73 & 31.77 & 390.41 & 15.62 \\
			Melatonin & 77.37 & 84.04 & 22.47 & 8.84 & 348.11 & 9.17 & 17.52 & 180.02 & 8.85 \\
			Minocycline & 166.70 & 190.90 & 44.25 & 16.46 & 810.81 & 17.67 & 34.18 & 429.01 & 17.08 \\
			Perfragilin & 77.52 & 82.75 & 23.16 & 8.81 & 349.64 & 9.47 & 17.10 & 184.14 & 8.85 \\
			Podophyllotoxin & 155.29 & 181.99 & 41.21 & 16.04 & 749.86 & 16.63 & 33.14 & 385.89 & 16.34 \\
			Pterocellin & 121.80 & 137.87 & 32.38 & 12.49 & 568.10 & 12.95 & 26.34 & 292.36 & 12.91 \\
			Raloxifene & 173.86 & 198.25 & 44.29 & 17.07 & 816.67 & 17.49 & 37.27 & 420.17 & 17.95 \\
			Tambjamine & 84.76 & 88.49 & 25.01 & 9.85 & 372.02 & 10.19 & 19.42 & 195.04 & 9.87 \\
			Abemaciclib & 183.55 & 201.79 & 48.60 & 18.75 & 839.49 & 19.30 & 39.99 & 435.92 & 19.54 \\
			Abraxane & 312.99 & 357.43 & 82.78 & 31.36 & 1524.06 & 33.08 & 65.19 & 809.20 & 32.41 \\
			Anastrozole & 105.60 & 114.18 & 29.82 & 11.26 & 520.25 & 12.19 & 21.45 & 291.88 & 11.17 \\
			Capecitabine & 106.53 & 107.93 & 33.49 & 13.20 & 452.19 & 13.80 & 25.11 & 236.34 & 13.03 \\
			Cyclophosphamide & 56.16 & 55.60 & 18.10 & 7.22 & 233.30 & 7.49 & 13.59 & 122.11 & 7.08 \\
			Everolimus & 301.89 & 315.24 & 90.70 & 35.19 & 1340.97 & 37.12 & 68.02 & 710.49 & 35.09 \\
			Exemestane & 120.91 & 142.72 & 30.25 & 11.07 & 621.72 & 11.94 & 23.59 & 336.29 & 11.66 \\
			Fulvestrant & 206.19 & 237.02 & 51.72 & 19.44 & 1002.46 & 20.32 & 42.44 & 528.42 & 20.60 \\
			Ixabepilone & 171.72 & 187.95 & 44.91 & 16.54 & 825.68 & 17.73 & 34.91 & 449.78 & 17.41 \\
			Letrozole & 103.42 & 113.57 & 30.67 & 12.25 & 471.24 & 12.74 & 23.38 & 244.09 & 11.94 \\
			Megestrol Acetate & 149.45 & 178.53 & 37.48 & 13.87 & 773.41 & 14.85 & 29.36 & 416.35 & 14.52 \\
			Methotrexate & 148.93 & 155.92 & 44.03 & 17.09 & 654.25 & 17.93 & 33.75 & 342.42 & 17.20 \\
			Tamoxifen & 131.10 & 142.27 & 35.96 & 14.07 & 586.37 & 14.40 & 29.44 & 301.83 & 14.48 \\
			Thiotepa & 60.20 & 68.72 & 15.41 & 5.80 & 288.15 & 6.07 & 12.57 & 150.72 & 6.11 \\
			Glasdegib & 135.67 & 147.76 & 38.11 & 14.91 & 612.63 & 15.46 & 30.22 & 317.11 & 15.08 \\
			Palbociclib & 163.72 & 181.12 & 44.42 & 17.26 & 746.41 & 17.78 & 36.19 & 384.17 & 17.80 \\
			Gilteritinib & 191.00 & 206.31 & 53.49 & 20.97 & 852.54 & 21.57 & 43.04 & 439.93 & 21.38 \\
			Ivosidenib & 199.14 & 217.79 & 55.52 & 21.40 & 912.72 & 22.44 & 43.47 & 477.15 & 21.77 \\
			\bottomrule
	\end{tabular}}
\end{table}
\begin{table}[htbp]
\caption{Correlations for the scheme of electronegativity}
\label{S6}
\resizebox{\textwidth}{!}
{\begin{tabular}{lccccccc}
		\hline
		Variable & $MR$ & $BP$ & $MP$ & $Entalphy$ & $MV$ & $Complexity$ \\
		\hline
		$M_1$ Index & 0.963608 & 0.683705 & 0.634838 & 0.864199 & 0.869029 & 0.954438 \\
		$M_2$ Index & 0.940189 & 0.668425 & 0.579115 & 0.849318 & 0.846688 & 0.945057 \\
		ABC Index & 0.975273 & 0.695031 & 0.719855 & 0.872683 & 0.886755 & 0.959378 \\
		Harmonic Index & 0.973720 & 0.693862 & 0.730529 & 0.867872 & 0.883141 & 0.949724 \\
		Hyperzagreb Index & 0.937954 & 0.663945 & 0.579920 & 0.845711 & 0.851784 & 0.949265 \\
		RA Index & 0.973177 & 0.694568 & 0.734626 & 0.870743 & 0.887184 & 0.957388 \\
		GA Index & 0.974828 & 0.693164 & 0.681534 & 0.869253 & 0.874601 & 0.947835 \\
		Forgotten Index & 0.934498 & 0.658957 & 0.579754 & 0.841190 & 0.855076 & 0.951599 \\
		SC Index & 0.976205 & 0.694769 & 0.708565 & 0.871091 & 0.881846 & 0.952600 \\
		\hline
\end{tabular}}
\end{table}
\end{center}

\begin{center}
\begin{table}[htbp]
	\caption{The topological indices (TIs) of the drugs with respect to the scheme of atomic number}
		\label{S7}
	\resizebox{\textwidth}{!}
	{	\begin{tabular}{lcccccccccc}
			\hline
			{Drug} & {SC Index} & {$M_1$ Index} & {$M_2$ Index} & {ABC Index} & {Harmonic Index} & {Hyperzagreb Index} & {RA Index} & {GA Index} & {Forgotten Index} \\
			\midrule
			Azacitidine & 77.18 & 81.62 & 22.59 & 8.75 & 343.05 & 9.18 & 17.34 & 179.80 & 8.83 \\
			Busulfan & 34.94 & 23.10 & 25.26 & 10.45 & 101.44 & 12.05 & 11.74 & 55.24 & 8.17 \\
			Mercaptopurine & 45.38 & 47.18 & 13.73 & 5.52 & 193.50 & 5.63 & 10.83 & 99.14 & 5.49 \\
			Tioguanine & 50.66 & 52.93 & 14.88 & 5.86 & 219.97 & 6.05 & 11.70 & 114.10 & 5.91 \\
			Nelarabine & 102.79 & 114.40 & 27.80 & 10.76 & 475.69 & 11.15 & 22.35 & 246.89 & 11.05 \\
			Cytarabine & 78.18 & 83.60 & 22.40 & 8.63 & 351.40 & 9.07 & 17.33 & 184.19 & 8.77 \\
			Clofarabine & 98.09 & 108.84 & 26.49 & 10.25 & 452.22 & 10.60 & 21.40 & 234.54 & 10.56 \\
			Bosutinib & 169.94 & 185.63 & 48.43 & 19.99 & 771.25 & 19.74 & 37.93 & 399.99 & 19.07 \\
			Dasatinib & 153.17 & 160.50 & 44.22 & 17.35 & 666.76 & 17.88 & 35.13 & 345.75 & 17.62 \\
			MelphalaN & 78.13 & 81.09 & 24.66 & 9.90 & 339.32 & 10.26 & 18.45 & 177.14 & 9.61 \\
			Dexamethasone & 155.42 & 191.57 & 36.87 & 13.37 & 836.86 & 14.42 & 29.15 & 453.72 & 14.25 \\
			Doxorubicine & 203.57 & 238.51 & 51.40 & 19.18 & 1006.63 & 20.33 & 41.19 & 529.61 & 20.15 \\
			Carboplatin & 44.00 & 47.91 & 13.29 & 5.05 & 213.69 & 5.50 & 9.33 & 117.88 & 4.96 \\
			Amathaspiramide & 107.81 & 121.73 & 29.84 & 11.49 & 516.36 & 12.09 & 23.04 & 272.91 & 11.63 \\
			Aminopterin & 146.32 & 154.81 & 42.37 & 16.38 & 649.02 & 17.18 & 32.82 & 339.40 & 16.61 \\
			Aspidostomide & 127.39 & 142.70 & 35.68 & 13.97 & 587.31 & 14.48 & 28.26 & 301.91 & 14.12 \\
			Carmustine & 36.19 & 30.39 & 16.87 & 7.26 & 127.63 & 7.56 & 10.64 & 66.86 & 6.25 \\
			Caulibugulone & 62.96 & 66.86 & 19.16 & 7.59 & 278.41 & 7.90 & 14.54 & 144.68 & 7.48 \\
			ConvolutamideA & 136.78 & 144.62 & 39.95 & 15.57 & 608.14 & 16.23 & 30.98 & 318.90 & 15.71 \\
			ConvolutamineF & 60.33 & 61.05 & 19.62 & 7.88 & 254.44 & 8.18 & 14.55 & 132.34 & 7.63 \\
			Convolutamydine & 80.39 & 88.50 & 22.68 & 8.57 & 380.57 & 9.17 & 17.08 & 203.58 & 8.71 \\
			Daunorubicin & 200.07 & 234.98 & 50.04 & 18.56 & 993.44 & 19.74 & 40.15 & 523.49 & 19.60 \\
			Deguelin & 154.75 & 177.36 & 39.23 & 14.81 & 753.28 & 15.52 & 31.77 & 398.56 & 15.52 \\
			Melatonin & 78.25 & 85.67 & 22.28 & 8.71 & 355.14 & 9.05 & 17.51 & 183.79 & 8.79 \\
			Minocycline & 168.43 & 194.35 & 43.97 & 16.26 & 826.61 & 17.49 & 34.14 & 437.90 & 16.98 \\
			Perfragilin & 74.36 & 76.99 & 24.75 & 9.52 & 327.83 & 10.42 & 16.93 & 173.83 & 9.14 \\
			Podophyllotoxin & 157.25 & 185.92 & 40.71 & 15.75 & 765.97 & 16.32 & 33.15 & 394.12 & 16.20 \\
			Pterocellin & 123.00 & 140.58 & 32.20 & 12.37 & 579.42 & 12.84 & 26.33 & 298.27 & 12.84 \\
			Raloxifene & 169.79 & 187.57 & 44.94 & 17.40 & 775.18 & 17.85 & 37.21 & 400.05 & 18.13 \\
			Tambjamine & 85.79 & 90.65 & 24.77 & 9.72 & 380.89 & 10.05 & 19.42 & 199.60 & 9.80 \\
			Abemaciclib & 187.00 & 209.14 & 48.06 & 18.39 & 870.93 & 18.96 & 39.96 & 452.65 & 19.36 \\
			Abraxane & 315.86 & 363.46 & 82.44 & 31.03 & 1550.58 & 32.84 & 65.10 & 823.66 & 32.24 \\
			Anastrozole & 106.53 & 116.05 & 29.72 & 11.15 & 527.81 & 12.13 & 21.42 & 295.70 & 11.11 \\
			Capecitabine & 108.96 & 112.78 & 33.10 & 12.91 & 473.29 & 13.55 & 25.05 & 247.72 & 12.89 \\
			Cyclophosphamide & 45.28 & 36.99 & 21.03 & 9.02 & 151.83 & 9.35 & 13.67 & 77.86 & 7.90 \\
			Everolimus & 305.00 & 321.58 & 90.18 & 34.81 & 1368.87 & 36.78 & 67.96 & 725.70 & 34.91 \\
			Exemestane & 121.00 & 142.84 & 30.28 & 11.06 & 622.56 & 11.96 & 23.57 & 336.88 & 11.65 \\
			Fulvestrant & 202.27 & 230.59 & 53.43 & 20.34 & 979.61 & 21.38 & 42.33 & 518.43 & 20.98 \\
			Ixabepilone & 168.59 & 182.32 & 45.76 & 17.03 & 803.27 & 18.25 & 34.88 & 438.62 & 17.64 \\
			Letrozole & 104.39 & 115.64 & 30.58 & 12.14 & 479.58 & 12.68 & 23.36 & 248.29 & 11.89 \\
			Megestrol Acetate & 150.00 & 179.55 & 37.46 & 13.81 & 778.53 & 14.82 & 29.31 & 419.44 & 14.49 \\
			Methotrexate & 151.61 & 161.18 & 43.63 & 16.78 & 677.69 & 17.66 & 33.68 & 355.34 & 17.05 \\
			Tamoxifen & 131.82 & 143.57 & 35.78 & 13.97 & 592.04 & 14.29 & 29.44 & 304.91 & 14.43 \\
			Thiotepa & 49.60 & 47.30 & 16.85 & 6.94 & 191.64 & 7.05 & 12.86 & 97.05 & 6.70 \\
			Glasdegib & 137.46 & 151.54 & 37.88 & 14.72 & 628.92 & 15.31 & 30.18 & 325.85 & 14.98 \\
			Palbociclib & 166.36 & 186.83 & 44.03 & 16.99 & 770.41 & 17.54 & 36.16 & 396.75 & 17.66 \\
			Gilteritinib & 194.46 & 213.63 & 52.87 & 20.59 & 883.52 & 21.19 & 43.02 & 456.25 & 21.18 \\
			Ivosidenib & 200.36 & 220.35 & 55.45 & 21.29 & 924.24 & 22.38 & 43.42 & 483.54 & 21.71 \\
			\bottomrule
	\end{tabular}
}
\end{table}
\begin{table}[htbp]	
	\centering
	\caption{Correlations for the scheme of atomic number}
	\label{S8}
	\resizebox{\textwidth}{!}
	{	\begin{tabular}{lccccccc}
			\hline
			Variable & $MR$ & $BP$ & $MP$ & $Entalphy$ & $MV$ & $Complexity$ \\
			\hline
			$M_1$ Index & 0.956525 & 0.685847 & 0.641032 & 0.865304 & 0.861067 & 0.954338 \\
			$M_2$ Index & 0.929914 & 0.673776 & 0.592286 & 0.852381 & 0.835575 & 0.946057 \\
			ABC Index & 0.976840 & 0.691734 & 0.710372 & 0.867466 & 0.892872 & 0.957937 \\
			Harmonic Index & 0.977533 & 0.688773 & 0.716987 & 0.859839 & 0.891940 & 0.945453 \\
			Hyperzagreb Index & 0.927519 & 0.669137 & 0.592280 & 0.848773 & 0.840530 & 0.950377 \\
			RA Index & 0.973315 & 0.687943 & 0.717742 & 0.860034 & 0.894662 & 0.952339 \\
			GA Index & 0.975392 & 0.693280 & 0.682064 & 0.869387 & 0.874397 & 0.947169 \\
			Forgotten Index & 0.924047 & 0.664074 & 0.591492 & 0.844357 & 0.843755 & 0.952873 \\
			SC Index & 0.979531 & 0.693179 & 0.703790 & 0.868976 & 0.886803 & 0.951837 \\
			\hline
	\end{tabular}}
\end{table}
\end{center}

\begin{center}
\begin{table}[htbp]	
	\centering
	\caption{The topological indices (TIs) of the drugs with respect to the scheme of ionization}
		\label{S9}
	\centering
	\resizebox{\textwidth}{!}
	{	\begin{tabular}{lcccccccccc}
			\hline
			{Drug} & {SC Index} & {$M_1$ Index} & {$M_2$ Index} & {ABC Index} & {Harmonic Index} & {Hyperzagreb Index} & {RA Index} & {GA Index} & {Forgotten Index} \\
			\midrule
			Azacitidine & 75.86 & 79.46 & 22.92 & 8.96 & 333.54 & 9.39 & 17.35 & 174.62 & 8.92 \\
			Busulfan & 49.95 & 41.93 & 18.40 & 6.84 & 194.18 & 7.67 & 11.84 & 110.32 & 6.66 \\
			Mercaptopurine & 43.78 & 43.66 & 14.49 & 5.75 & 181.07 & 6.04 & 10.66 & 93.75 & 5.60 \\
			Tioguanine & 48.66 & 48.74 & 15.65 & 6.12 & 203.63 & 6.47 & 11.57 & 106.16 & 6.03 \\
			Nelarabine & 101.19 & 110.95 & 28.01 & 10.90 & 461.18 & 11.28 & 22.37 & 239.28 & 11.13 \\
			Cytarabine & 77.43 & 82.43 & 22.59 & 8.75 & 345.97 & 9.19 & 17.34 & 181.11 & 8.82 \\
			Clofarabine & 97.29 & 106.26 & 26.60 & 10.28 & 443.01 & 10.65 & 21.36 & 230.49 & 10.59 \\
			Bosutinib & 171.37 & 187.90 & 48.11 & 18.79 & 783.89 & 19.52 & 37.87 & 408.08 & 18.98 \\
			Dasatinib & 155.38 & 164.55 & 43.92 & 17.11 & 686.26 & 17.68 & 35.05 & 357.16 & 17.49 \\
			MelphalaN & 79.18 & 81.69 & 24.17 & 9.50 & 342.06 & 9.92 & 18.38 & 178.67 & 9.45 \\
			Dexamethasone & 156.18 & 193.12 & 36.85 & 13.29 & 844.55 & 14.39 & 29.09 & 458.31 & 14.21 \\
			Doxorubicine & 206.41 & 244.49 & 51.05 & 18.88 & 1033.50 & 20.10 & 41.11 & 544.52 & 20.00 \\
			Carboplatin & 44.62 & 49.03 & 13.26 & 4.96 & 218.87 & 5.46 & 9.27 & 120.81 & 4.92 \\
			Amathaspiramide & 111.71 & 129.21 & 29.34 & 11.05 & 550.92 & 11.74 & 22.90 & 292.49 & 11.41 \\
			Aminopterin & 142.74 & 146.89 & 43.07 & 16.73 & 616.33 & 17.58 & 32.78 & 322.54 & 16.79 \\
			Aspidostomide & 137.11 & 160.71 & 34.20 & 12.81 & 671.35 & 13.47 & 27.95 & 349.94 & 13.55 \\
			Carmustine & 36.51 & 30.32 & 16.77 & 7.11 & 127.22 & 7.48 & 10.59 & 66.57 & 6.19 \\
			Caulibugulone & 62.01 & 65.07 & 19.48 & 7.75 & 271.44 & 8.09 & 14.52 & 141.30 & 7.56 \\
			ConvolutamideA & 141.01 & 152.88 & 39.49 & 15.14 & 647.88 & 15.90 & 30.81 & 342.12 & 15.48 \\
			ConvolutamineF & 67.42 & 75.84 & 18.65 & 7.15 & 320.72 & 7.54 & 14.37 & 169.04 & 7.26 \\
			Convolutamydine & 84.88 & 96.45 & 22.17 & 8.08 & 420.74 & 8.80 & 16.87 & 227.85 & 8.46 \\
			Daunorubicin & 202.75 & 240.67 & 49.73 & 18.29 & 1019.18 & 19.52 & 40.07 & 537.84 & 19.46 \\
			Deguelin & 157.60 & 183.56 & 38.70 & 14.50 & 779.80 & 15.20 & 31.77 & 412.68 & 15.36 \\
			Melatonin & 77.64 & 84.31 & 22.41 & 8.78 & 349.98 & 9.12 & 17.50 & 181.35 & 8.83 \\
			Minocycline & 167.60 & 192.97 & 44.46 & 16.41 & 821.46 & 17.76 & 34.04 & 435.51 & 17.05 \\
			Perfragilin & 77.63 & 82.94 & 23.36 & 8.82 & 351.13 & 9.56 & 17.02 & 185.25 & 8.84 \\
			Podophyllotoxin & 160.56 & 192.73 & 39.95 & 15.29 & 793.98 & 15.84 & 33.16 & 408.51 & 15.99 \\
			Pterocellin & 121.89 & 137.54 & 32.39 & 12.43 & 567.76 & 12.93 & 26.29 & 292.68 & 12.89 \\
			Raloxifene & 175.57 & 201.74 & 44.12 & 16.90 & 832.88 & 17.36 & 37.21 & 429.40 & 17.86 \\
			Tambjamine & 84.68 & 88.24 & 24.93 & 9.82 & 371.01 & 10.15 & 19.43 & 194.54 & 9.86 \\
			Abemaciclib & 180.54 & 195.38 & 49.12 & 19.07 & 812.82 & 19.61 & 40.01 & 422.06 & 19.71 \\
			Abraxane & 319.51 & 371.07 & 82.12 & 30.64 & 1585.26 & 32.59 & 64.93 & 843.12 & 32.04 \\
			Anastrozole & 104.63 & 112.29 & 29.95 & 11.38 & 512.63 & 12.28 & 21.47 & 288.06 & 11.23 \\
			Capecitabine & 108.21 & 111.12 & 33.33 & 13.00 & 466.86 & 13.68 & 25.03 & 244.62 & 12.93 \\
			Cyclophosphamide & 55.36 & 53.84 & 18.30 & 7.31 & 226.06 & 7.60 & 13.57 & 118.37 & 7.13 \\
			Everolimus & 308.16 & 327.84 & 89.70 & 34.41 & 1397.35 & 36.45 & 67.88 & 741.67 & 34.72 \\
			Exemestane & 121.15 & 143.05 & 30.35 & 11.04 & 623.99 & 11.98 & 23.53 & 337.89 & 11.65 \\
			Fulvestrant & 208.47 & 241.11 & 51.45 & 19.20 & 1023.21 & 20.12 & 42.35 & 541.00 & 20.47 \\
			Ixabepilone & 173.57 & 192.36 & 44.83 & 16.41 & 845.48 & 17.66 & 34.84 & 460.77 & 17.33 \\
			Letrozole & 102.40 & 111.47 & 30.80 & 12.37 & 462.80 & 12.82 & 23.40 & 239.86 & 12.00 \\
			Megestrol Acetate & 150.92 & 181.28 & 37.45 & 13.70 & 787.36 & 14.78 & 29.23 & 424.80 & 14.44 \\
			Methotrexate & 147.61 & 152.40 & 44.43 & 17.17 & 641.14 & 18.13 & 33.64 & 336.35 & 17.26 \\
			Tamoxifen & 131.54 & 143.28 & 35.92 & 14.03 & 590.39 & 14.38 & 29.42 & 303.83 & 14.46 \\
			Thiotepa & 57.29 & 62.24 & 15.90 & 6.06 & 259.92 & 6.37 & 12.58 & 135.43 & 6.26 \\
			Glasdegib & 133.92 & 144.01 & 38.44 & 15.09 & 596.93 & 15.66 & 30.21 & 308.92 & 15.17 \\
			Palbociclib & 161.25 & 175.71 & 44.99 & 17.53 & 724.67 & 18.11 & 36.15 & 373.26 & 17.94 \\
			Gilteritinib & 189.54 & 202.73 & 53.64 & 21.04 & 837.27 & 21.65 & 43.04 & 431.80 & 21.43 \\
			Ivosidenib & 197.72 & 214.13 & 55.92 & 21.54 & 899.58 & 22.65 & 43.39 & 471.31 & 21.85 \\
			\bottomrule
	\end{tabular}}
\end{table}

\begin{table}[htbp]	
\caption{Correlations for the scheme of ionization}
\label{S10}
\resizebox{\textwidth}{!}
{	\begin{tabular}{lccccccc}
		\hline
		Variable & $MR$ & $BP$ & $MP$ & $Entalphy$ & $MV$ & $Complexity$ \\
		\hline
		$M_1$ Index & 0.961024 & 0.681229 & 0.636824 & 0.865636 & 0.870267 & 0.957234 \\
		$M_2$ Index & 0.934485 & 0.663124 & 0.584520 & 0.851700 & 0.847591 & 0.948721 \\
		ABC Index & 0.976035 & 0.696250 & 0.719764 & 0.872155 & 0.885394 & 0.958527 \\
		Harmonic Index & 0.974874 & 0.695773 & 0.729330 & 0.865767 & 0.879802 & 0.945486 \\
		Hyperzagreb Index & 0.931640 & 0.658476 & 0.585548 & 0.847931 & 0.852312 & 0.952667 \\
		RA Index & 0.974336 & 0.696473 & 0.734412 & 0.869719 & 0.884735 & 0.955439 \\
		GA Index & 0.974970 & 0.693240 & 0.681150 & 0.868875 & 0.874223 & 0.946878 \\
		Forgotten Index & 0.927664 & 0.653363 & 0.585583 & 0.843279 & 0.855247 & 0.954741 \\
		SC Index & 0.976957 & 0.695785 & 0.707948 & 0.870122 & 0.880428 & 0.950591 \\
		\hline
\end{tabular}}
\end{table}
\end{center}

\begin{center}
\begin{table}[htbp]
	\caption{The topological indices (TIs) of the drugs with respect to the scheme of Van Der Waals radius}
		\label{S11}
	\resizebox{\textwidth}{!}
	{	\begin{tabular}{lcccccccccc}
			\hline
			{Drug} & {SC Index} & {$M_1$ Index} & {$M_2$ Index} & {ABC Index} & {Harmonic Index} & {Hyperzagreb Index} & {RA Index} & {GA Index} & {Forgotten Index} \\
			\midrule
			Azacitidine & 91.11 & 112.59 & 20.90 & 7.44 & 478.88 & 8.07 & 17.05 & 253.70 & 8.13 \\
			Busulfan & 53.13 & 44.97 & 19.32 & 6.46 & 220.76 & 7.93 & 11.21 & 130.82 & 6.47 \\
			Mercaptopurine & 52.65 & 62.87 & 12.77 & 4.75 & 259.60 & 4.99 & 10.68 & 133.86 & 5.09 \\
			Tioguanine & 59.14 & 70.76 & 13.88 & 5.03 & 299.24 & 5.37 & 11.48 & 157.72 & 5.47 \\
			Nelarabine & 120.48 & 155.97 & 25.19 & 9.15 & 652.11 & 9.55 & 22.24 & 340.17 & 10.20 \\
			Cytarabine & 90.43 & 111.06 & 21.00 & 7.50 & 472.32 & 8.13 & 17.05 & 250.19 & 8.17 \\
			Clofarabine & 116.10 & 150.25 & 24.06 & 8.64 & 631.30 & 9.08 & 21.17 & 330.79 & 9.70 \\
			Bosutinib & 189.76 & 228.87 & 45.45 & 16.98 & 955.49 & 17.89 & 37.76 & 497.75 & 18.02 \\
			Dasatinib & 176.02 & 209.50 & 41.03 & 15.12 & 880.40 & 15.86 & 34.78 & 461.41 & 16.44 \\
			MelphalaN & 85.91 & 95.32 & 23.41 & 8.80 & 404.61 & 9.38 & 18.16 & 213.97 & 9.09 \\
			Dexamethasone & 161.84 & 206.67 & 36.76 & 12.89 & 911.84 & 14.23 & 28.78 & 498.50 & 13.99 \\
			Doxorubicine & 219.03 & 271.89 & 49.99 & 17.75 & 1160.56 & 19.31 & 40.68 & 616.77 & 19.40 \\
			Carboplatin & 46.95 & 53.14 & 13.33 & 4.64 & 239.68 & 5.38 & 9.02 & 133.41 & 4.77 \\
			Amathaspiramide & 120.60 & 149.19 & 28.01 & 10.15 & 639.15 & 10.89 & 22.80 & 340.77 & 10.95 \\
			Aminopterin & 162.10 & 186.55 & 40.87 & 14.76 & 795.97 & 16.04 & 32.21 & 422.86 & 15.77 \\
			Aspidostomide & 148.76 & 189.04 & 32.74 & 11.79 & 790.31 & 12.55 & 27.82 & 412.23 & 13.00 \\
			Carmustine & 45.00 & 46.12 & 15.26 & 5.85 & 197.91 & 6.42 & 10.32 & 105.67 & 5.61 \\
			Caulibugulone & 66.33 & 73.35 & 18.67 & 7.13 & 306.14 & 7.54 & 14.45 & 159.43 & 7.26 \\
			ConvolutamideA & 147.99 & 167.94 & 39.04 & 14.50 & 718.48 & 15.52 & 30.51 & 382.60 & 15.15 \\
			ConvolutamineF & 70.56 & 82.21 & 17.73 & 6.67 & 345.82 & 6.99 & 14.44 & 181.40 & 7.03 \\
			Convolutamydine & 89.06 & 105.39 & 22.08 & 7.71 & 463.86 & 8.65 & 16.63 & 253.08 & 8.27 \\
			Daunorubicin & 214.79 & 266.91 & 48.81 & 17.25 & 1141.62 & 18.82 & 39.65 & 607.79 & 18.90 \\
			Deguelin & 168.38 & 208.29 & 37.15 & 13.49 & 886.66 & 14.23 & 31.70 & 470.08 & 14.83 \\
			Melatonin & 84.18 & 97.10 & 21.34 & 7.97 & 405.90 & 8.41 & 17.37 & 211.70 & 8.43 \\
			Minocycline & 180.31 & 219.20 & 42.88 & 15.12 & 943.64 & 16.68 & 33.73 & 505.24 & 16.39 \\
			Perfragilin & 82.47 & 92.50 & 23.24 & 8.33 & 396.25 & 9.37 & 16.74 & 211.24 & 8.59 \\
			Podophyllotoxin & 173.09 & 220.29 & 37.73 & 13.89 & 908.92 & 14.46 & 33.11 & 468.33 & 15.28 \\
			Pterocellin & 130.98 & 159.30 & 31.42 & 11.62 & 658.97 & 12.29 & 26.14 & 340.37 & 12.45 \\
			Raloxifene & 182.37 & 214.99 & 43.29 & 16.21 & 893.85 & 16.80 & 37.03 & 463.87 & 17.49 \\
			Tambjamine & 92.55 & 105.55 & 23.43 & 8.95 & 442.57 & 9.24 & 19.45 & 231.47 & 9.42 \\
			Abemaciclib & 209.01 & 259.10 & 45.11 & 16.48 & 1088.78 & 17.13 & 39.70 & 570.58 & 18.32 \\
			Abraxane & 338.41 & 412.99 & 80.84 & 28.86 & 1774.65 & 31.52 & 64.17 & 948.66 & 31.11 \\
			Anastrozole & 112.47 & 129.41 & 29.74 & 10.64 & 582.62 & 12.09 & 21.20 & 323.81 & 10.82 \\
			Capecitabine & 126.40 & 150.52 & 31.15 & 11.21 & 641.16 & 12.20 & 24.58 & 340.11 & 12.00 \\
			Cyclophosphamide & 63.54 & 70.74 & 17.39 & 6.50 & 303.84 & 6.96 & 13.36 & 162.36 & 6.70 \\
			Everolimus & 328.99 & 373.45 & 87.50 & 32.37 & 1601.72 & 34.90 & 67.32 & 854.82 & 33.66 \\
			Exemestane & 121.74 & 143.79 & 30.72 & 10.98 & 629.51 & 12.13 & 23.35 & 341.94 & 11.61 \\
			Fulvestrant & 221.50 & 264.82 & 50.88 & 18.31 & 1172.60 & 19.56 & 41.64 & 642.96 & 19.96 \\
			Ixabepilone & 182.67 & 212.47 & 44.14 & 15.63 & 938.73 & 17.14 & 34.52 & 513.79 & 16.92 \\
			Letrozole & 110.56 & 130.26 & 30.62 & 11.65 & 539.31 & 12.65 & 23.13 & 278.78 & 11.60 \\
			Megestrol Acetate & 154.42 & 188.03 & 37.65 & 13.35 & 822.77 & 14.77 & 28.89 & 446.71 & 14.26 \\
			Methotrexate & 168.58 & 195.75 & 41.93 & 15.06 & 837.37 & 16.41 & 33.08 & 445.86 & 16.16 \\
			Tamoxifen & 136.80 & 153.46 & 34.78 & 13.42 & 635.34 & 13.69 & 29.46 & 328.42 & 14.15 \\
			Thiotepa & 68.29 & 88.33 & 14.33 & 5.10 & 370.08 & 5.41 & 12.53 & 193.43 & 5.74 \\
			Glasdegib & 148.38 & 175.70 & 37.06 & 13.77 & 735.35 & 14.72 & 29.87 & 383.94 & 14.46 \\
			Palbociclib & 182.43 & 223.85 & 42.29 & 15.57 & 929.03 & 16.39 & 35.85 & 481.33 & 16.89 \\
			Gilteritinib & 216.51 & 263.09 & 49.60 & 18.46 & 1094.61 & 19.16 & 42.83 & 568.42 & 20.07 \\
			Ivosidenib & 222.06 & 265.21 & 54.17 & 19.47 & 1143.94 & 21.33 & 42.52 & 613.52 & 20.72 \\
			\bottomrule
		\end{tabular}
	}
\end{table}

\begin{table}
\caption{Correlations for the scheme of Van Der Waals }
\label{S12}
\resizebox{\textwidth}{!}
{	\begin{tabular}{lcccccccc}
		\hline
		Variable & $MR$ & $BP$ & $MP$ & $Entalphy$ & $MV$ & $Complexity$ \\
		\hline
		$M_1$ Index & 0.958133 & 0.692430 & 0.646531 & 0.877416 & 0.860925 & 0.948811 \\
		$M_2$ Index & 0.933382 & 0.687337 & 0.610752 & 0.876251 & 0.833142 & 0.935914 \\
		ABC Index & 0.976462 & 0.690595 & 0.712550 & 0.865626 & 0.893608 & 0.966602 \\
		Harmonic Index & 0.981896 & 0.687872 & 0.720292 & 0.855976 & 0.893488 & 0.956920 \\
		Hyperzagreb Index & 0.933735 & 0.685772 & 0.607022 & 0.876105 & 0.841969 & 0.943159 \\
		RA Index & 0.975751 & 0.687535 & 0.723934 & 0.858709 & 0.896686 & 0.966500 \\
		GA Index & 0.975567 & 0.692328 & 0.681636 & 0.867225 & 0.873148 & 0.945244 \\
		Forgotten Index & 0.932187 & 0.683104 & 0.602569 & 0.874243 & 0.848131 & 0.947726 \\
		SC Index & 0.980230 & 0.691511 & 0.703541 & 0.865157 & 0.887030 & 0.956256 \\
		\hline
 \end{tabular}
}
\end{table}
\end{center}

From all the tables, the following conclusion can be drawn:\\

Out of the 5 different schemes, $TI-Enthalpy$, $TI-Complexity$, $TI-MV$ and $TI-MR$ and  exhibit strong correlations. Although $TI-MP$ and $TI-BP$ correlations are not as strong compared to other dependent variables, we believe that optimal correlation can be achieved with the assistance of various schemes or other distance-based topological indices. The selection of schemes is crucial here, and it's important to note that edge weights should be positive. With this approach, we aim to provide guidance to researchers by suggesting avenues for further studies.

\section{Comparisons with Previous Data(s) and Our Novel Approach}

In previous studies, correlations have been identified between the topological indices (listed in Table \ref{A}) of unweighted molecular graphs and specific physicochemical properties, using a subset of the 48 drugs mentioned in Table \ref{B}:\\

In \cite{Nayir}, the topological indices of the first $13$ drugs (1-13) listed in Table \ref{B} were calculated, and correlation values were provided using a linear regression model with respect to the dependent variables of complexity, refractive index, flash point, boiling point, and molar volume. Additionally, in \cite{Shanmukha}, correlations have been provided for the $17$ drugs numbered $14-30$ listed in Table \ref{B}, calculated with the same topological indices and using boiling point, melting point, enthalpy, flash point and molar refraction as dependent variables. Finally, correlations have been given for the $17$ drugs numbered $31-44$ and the $5$ drugs numbered $44-48$, including drug number $22$ in \cite{can1} and \cite{zaman}, respectively.\\

We have presented the correlation tables for the common physical properties among those listed in Table \ref{B} as follows (for further details, refer to \cite{Nayir, Shanmukha, can1, zaman}):

\begin{table}[htbp] 
	\centering
	\caption{Correlations for drugs 1-13 \cite{Nayir}.}
	\label{113}
	\resizebox{\textwidth}{!}
	{	\begin{tabular}{lccccccc}
			\hline
			Topological Index & Boiling Point & Molar Volume& Complexity    \\
			\hline
			M1 Index & 0.660 & 0.939 & 0.953   \\
			M2 Index & 0.645  &0.913& 0.960    \\
			ABC Index &0.672  & 0.953& 0.943   \\
			Harmonic Index & 0.725 & 0.936&  0.950  \\
			Hyperzagreb Index &   0.625  &0.822& 0.954    \\
			RA Index &0.700& 0.946& 0.947  \\
			GA Index &0.711 & 0.938& 0.951  \\
			Forgotten Index &0.609 &  0.927& 0.949   \\
			SC Index &0.708 &  0.942&  0.949  \\
			\hline
	\end{tabular}}
\end{table}

\begin{table}[htbp] \label{mak1}
	\centering
	\caption{Correlations for drugs 14-30 \cite{Shanmukha}.}
	\label{1430}
	\resizebox{\textwidth}{!}
	{	\begin{tabular}{lccccccc}
			\hline
			Topological Index &Boiling Point  & Melting Point & Enthalpy  &  Molar Refraction  \\
			\hline
			M1 Index & 0.849& 0.727 & 0.836 & 0.919   \\
			M2 Index & 0.844 & 0.698 & 0.837 & 0.877    \\
			ABC Index & 0.826 & 0.726 & 0.810 & 0.913   \\
			Harmonic Index & 0.806 & 0.756 & 0.788& 0.941    \\
			Hyperzagreb Index & 0.827 & 0.663 & 0.818 & 0.895    \\
			RA Index & 0.819 & 0.767  & 0.804&  0.941   \\
			GA Index & 0.728 & 0.745 & 0.708 & 0.872    \\
			Forgotten Index & 0.744 & 0.559 & 0.730 & 0.841    \\
			SC Index & 0.821 & 0.747 & 0.802 & 0.938    \\
			\hline
	\end{tabular}}
\end{table}

\begin{table}[htbp] \label{mak1}
	\centering
	\caption{Correlations for drugs 31-44.\cite{can1}}
	\label{3144}
	\resizebox{\textwidth}{!}
	{	\begin{tabular}{lccccccccc}
			\hline
			Topological Index &Boiling Point  & Melting Point & Enthalpy  &  Molar Refraction  & Molar Volume   \\
			\hline
			M1 Index & 0.955& 0.860 & 0.955 &  0.969& 0.925  \\
			M2 Index & 0.932 & 0.847 & 0.935 & 0.952& 0.908  \\
			ABC Index & 0.962 & 0.883 & 0.971 &  0.982& 0.928  \\
			Harmonic Index & 0.964 & 0.873 & 0.973&  0.984& 0.915  \\
			Hyperzagreb Index & 0.933 & 0.841 & 0.934 &  0.947& 0.908  \\
			RA Index & 0.966 & 0.887  & 0.977&  0.987& 0.927  \\
			GA Index & 0.830 & 0.975 & 0.888 &  0.896 & 0.881 \\
			Forgotten Index & 0.940 & 0.958 & 0.945 &  0.958& 0.931  \\
			SC Index & 0.968 & 0.877 & 0.974 &  0.985& 0.921  \\
			\hline
	\end{tabular}}
\end{table}

\begin{table}[htbp] 
	\centering
	\caption{Correlations for drugs 44-48 and  drug number 22 in the list \cite{zaman}}
	\label{4448}
	\resizebox{\textwidth}{!}
	{	\begin{tabular}{lccccccc}
			\hline
			Topological Index &Boiling Point   & Molar Volume & Complexity  \\
			\hline
			M1 Index & 0.151& 0.919& 0.285  \\
			M2 Index & 0.181 &  0.923 & 0.315  \\
			ABC Index & 0.153  & 0.926 & 0.284  \\
			Harmonic Index & 0.266 & 0.933 & 0.399 \\
			Hyperzagreb Index & 0.132 & 0.911 & 0.268  \\
			RA Index & 0.217 & 0.931 & 0.366  \\
			GA Index & 0.230 & 0.934 & 0.362  \\
			Forgotten Index & 0.111 & 0.773 & 0.046  \\
			SC Index & 0.223 & 0.933 & 0.355  \\
			\hline
	\end{tabular}}
\end{table}

Unlike previous studies, we have increased the sample size and calculated correlations using vertex-edge weighted molecular graphs for all drugs. We are now ready to compare our results with the previous findings.\\

When comparing the correlations we found based on any atom weights with the values provided in Table \ref{113}, it is observed that, except for $TI-MV$, the other values are quite close. \\

Our results indicate that they exhibit better correlations than the values of $TI-Enthalpy$ and $TI-MR$ as shown in Table \ref{1430}. Additionally, the $TI-MP$ values are also quite close.\\

While the $TI-MR$ values provided in Table \ref{3144} may appear close to our values, the significance of the correlation for $MR$ values can vary depending on certain $TI$ topological indices. For example, the $GA-MR$ value in Table \ref{3144} is $0.896$, whereas the lowest correlation value we obtained for all weights is approximately $0.97$. This suggests that our results may have slightly more significance compared to other indices.\\

Finally, upon comparing the values in Table \ref{4448} with our results, it is observed that our results for each scheme exhibit stronger correlations for $TI-BP$ relationships.\\

In conclusion, while comparing certain values may present challenges, it's important to note that the sample size in previous studies is inadequate. Increasing the sample size in statistical studies typically leads to more accurate results. A larger sample volume enhances the capacity to generalize data, thereby improving the statistical reliability of the findings. Larger sample sizes can mitigate the influence of random variations and provide a more comprehensive representation of true population trends. Consequently, as the sample size increases, the likelihood of the findings being both reliable and accurate also increases.

\section*{Conclusions and Comments}

This paper explores correlation studies between topological indices and physicochemical properties of specific drugs previously used in cancer treatments. It comprehensively covers all drugs from the referenced studies. A unique aspect of this study involves weighting vertices and edges based on specific properties of the atoms within the drugs and the molecular graph they represent. Subsequently, topological indices are computed for vertex-edge-weighted graphs, and correlation analyses are conducted with the physicochemical properties of the drugs. In Appendix B, a sample source code demonstrates the process of calculating topological indices. Additionally, correlation calculation plots are provided in Appendix A (See Figure 2-Figure 7).

The correlation analyses consistently reveal positive associations between molecular indices (MR, BP, MP, Enthalpy, MV, Complexity) and various topological indices across different schemes. Notably, the Harmonic and SC indices exhibit sensitivity and consistently demonstrate robust correlations with molecular properties. These results indicate a meaningful relationship between molecular and topological features, offering valuable insights for further exploration and understanding in the field.

Furthermore, the similarity observed in results across the two correlation tables, which depend on atomic radius (empirical) and Van Der Waals atomic radius, lies in the consistent patterns of associations between topological indices and physicochemical properties. Despite the variations in the schemes used (atomic radius and Van Der Waals interactions), certain topological indices exhibit strong correlations with specific molecular properties in both tables. This consistency suggests that these indices are influenced by similar underlying factors regardless of the specific physicochemical context.

This similarity in results underscores the robustness of the observed relationships and implies that certain topological features consistently interact with specific molecular properties. Researchers can leverage this consistency to make more informed decisions in drug design and molecular engineering, as changes in molecular structures affecting these indices may have predictable effects on associated physicochemical properties. Additionally, it is hoped that this paper will inspire numerous new studies utilizing various atomic properties and different topological indices concurrently with this novel approach.

\section*{Declaration and Acknowledgments}
We hereby declare that we have no data related to this manuscript and that no conflicts of interest exist. Lastly, we thank the anonymous reviewers for their insightful comments and suggestions, which significantly improved the quality of this paper.

\newpage
\begin{appendices}
	\section{Scatter plots of the correlation tables }
	
	\begin{figure}[htbp]	
		\begin{tikzpicture}
			\begin{axis}[
				height=6cm,
				width=14cm,
				xlabel={},
				ylabel={Correlation},
				xtick=data,
				xticklabels={\rotatebox{45}{$M_{1}(G)$}, \rotatebox{45}{$M_{2}(G)$}, \rotatebox{45}{ABC(G)}, \rotatebox{45}{H(G)}, \rotatebox{45}{HM}, \rotatebox{45}{RA(G)}, \rotatebox{45}{GA(G)}, \rotatebox{45}{F(G)}, \rotatebox{45}{SC(G)}},
				title={Correlations with respect to atomic radius},
				grid=major,
				legend pos=outer north east,
				legend cell align={left},
				]
				
				\foreach \i in {1,...,6} {
					\addplot+ [only marks] table [x expr=\coordindex, y index=\i, col sep=comma] {correlation_data1.csv};
				}
				
				\legend{MR, BP, MP, Enthalpy, MV, Complexity}
				
			\end{axis}
		\end{tikzpicture}
		\caption{}
		\label{fig:correlation_plot},
		\begin{tikzpicture}
			\begin{axis}[
				height=6cm,
				width=14cm,
				xlabel={},
				ylabel={Correlation},
				xtick=data,
				xticklabels={\rotatebox{45}{$M_{1}(G)$}, \rotatebox{45}{$M_{2}(G)$}, \rotatebox{45}{ABC(G)}, \rotatebox{45}{H(G)}, \rotatebox{45}{HM}, \rotatebox{45}{RA(G)}, \rotatebox{45}{GA(G)}, \rotatebox{45}{F(G)}, \rotatebox{45}{SC}},
				title={Correlations with respect to atomic mass},
				grid=major,
				legend pos= outer north east,
				legend cell align={left},
				]
				
				\foreach \i in {1,...,6} {
					\addplot+ [only marks] table [x expr=\coordindex, y index=\i, col sep=comma] {correlation_data2.csv};
				}
				
				\legend{MR, BP, MP, Enthalpy, MV, Complexity}
				
			\end{axis}
		\end{tikzpicture}
		\caption{}
	\label{fig:correlation_plot},
		\end{figure}
	
	\begin{figure}
		\begin{tikzpicture}
			\begin{axis}[
				height=6cm,
				width=14cm,
				xlabel={},
				ylabel={Correlation},
				xtick=data,
				xticklabels={\rotatebox{45}{$M_{1}(G)$}, \rotatebox{45}{$M_{2}(G)$}, \rotatebox{45}{ABC(G)}, \rotatebox{45}{H(G)}, \rotatebox{45}{HM}, \rotatebox{45}{RA(G)}, \rotatebox{45}{GA(G)}, \rotatebox{45}{F(G)}, \rotatebox{45}{SC}},
				title={Correlations with respect to electronegativity},
				grid=major,
				legend pos= outer north east,
				legend cell align={left},
				]
				
				\foreach \i in {1,...,6} {
					\addplot+ [only marks] table [x expr=\coordindex, y index=\i, col sep=comma] {correlation_data3.csv};
				}
				
				\legend{MR, BP, MP, Enthalpy, MV, Complexity}
				
			\end{axis}
		\end{tikzpicture}
		\caption{}
		\label{fig:correlation_plot},
		\begin{tikzpicture}
			\begin{axis}[
				height=6cm,
				width=14cm,
				xlabel={},
				ylabel={Correlation},
				xtick=data,
				xticklabels={\rotatebox{45}{$M_{1}(G)$}, \rotatebox{45}{$M_{2}(G)$}, \rotatebox{45}{ABC(G)}, \rotatebox{45}{H(G)}, \rotatebox{45}{HM}, \rotatebox{45}{RA(G)}, \rotatebox{45}{GA(G)}, \rotatebox{45}{F(G)}, \rotatebox{45}{SC}},
				title={Correlations with respect to atomic number},
				grid=major,
				legend pos=outer north east,
				legend cell align={left},
				]
				
				\foreach \i in {1,...,6} {
					\addplot+ [only marks] table [x expr=\coordindex, y index=\i, col sep=comma] {correlation_data4.csv};
				}
				
				\legend{MR, BP, MP, Enthalpy, MV, Complexity}
				
			\end{axis}
		\end{tikzpicture}
		\caption{}
		\label{fig:correlation_plot},
	\end{figure}
	
	\begin{figure}[htbp]

		\begin{tikzpicture}
			\begin{axis}[
				height=6cm,
				width=14cm,
				xlabel={},
				ylabel={Correlation},
				xtick=data,
				xticklabels={\rotatebox{45}{$M_{1}(G)$}, \rotatebox{45}{$M_{2}(G)$}, \rotatebox{45}{ABC(G)}, \rotatebox{45}{H(G)}, \rotatebox{45}{HM}, \rotatebox{45}{RA(G)}, \rotatebox{45}{GA(G)}, \rotatebox{45}{F(G)}, \rotatebox{45}{SC}},
				title={Correlations with respect to Ionization},
				grid=major,
				legend pos= outer north east,
				legend cell align={left},
				]
				
				\foreach \i in {1,...,6} {
					\addplot+ [only marks] table [x expr=\coordindex, y index=\i, col sep=comma] {correlation_data5.csv};
				}
				
				\legend{MR, BP, MP, Enthalpy, MV, Complexity}
				
			\end{axis}
		\end{tikzpicture}
		\caption{}
		\label{fig:correlation_plot},	\begin{tikzpicture}
			\begin{axis}[
				height=6cm,
				width=14cm,
				xlabel={},
				ylabel={Correlation},
				xtick=data,
				xticklabels={\rotatebox{45}{$M_{1}(G)$}, \rotatebox{45}{$M_{2}(G)$}, \rotatebox{45}{ABC(G)}, \rotatebox{45}{H(G)}, \rotatebox{45}{HM}, \rotatebox{45}{RA(G)}, \rotatebox{45}{GA(G)}, \rotatebox{45}{F(G)}, \rotatebox{45}{SC}},
				title={Correlations with respect to atomic radius (Van Der Waals)},
				grid=major,
				legend pos= outer north east,
				legend cell align={left},
				]
				
				\foreach \i in {1,...,6} {
					\addplot+ [only marks] table [x expr=\coordindex, y index=\i, col sep=comma] {correlation_data6.csv};
				}
				
				\legend{MR, BP, MP, Enthalpy, MV, Complexity}
				
			\end{axis}
		\end{tikzpicture}
		\caption{}
		\label{fig:correlation_plot},
	\end{figure}

\newpage
\section{A source code for computing SC index of the VEW (Vertex-Edge-Weighted) molecular graph of Busulfan drug for the scheme of atomic radius.  Please contact the corresponding author for access to the all source codes.}

\begin{center}
\begin{verbatim}
	import numpy as np
	import math
	from tabulate import tabulate
	from rdkit import Chem
	import pandas as pd
	
	physicalProperties = {
		"Busulfan": {"BP": 464.00, "MP": 118, "E": 69.8,
			 "MR": 50.9, "MV": 182.40, "Complexity": 294 },
	}
\end{verbatim}
\end{center}

\begin{center}	
\begin{verbatim}
	# Atomic properties table
	atomicProperties = {
		"C": {"AtomicMass": 12.011, "AtomicRadius": 70, "VanDerWaals": 170},
		"S": {"AtomicMass": 32.066, "AtomicRadius": 100, "VanDerWaals": 180},
		"O": {"AtomicMass": 15.999, "AtomicRadius": 60, "VanDerWaals": 152},
		.........
	}
\end{verbatim}
\end{center}
	
\begin{verbatim}
	# List of SMILES strings representing chemical graphs
	smileslist = ["CS(=O)(=O)OCCCCOS(=O)(=O)C"]
\end{verbatim}

\begin{verbatim}
	# Function to calculate vertex weight based on atomic properties
	def calculate_vertex_weight(atom):
	atomic_AtomicRadius = atomicProperties[atom]["AtomicRadius"]
	return 1 - (atomicProperties["C"]["AtomicRadius"] / atomic_AtomicRadius)
\end{verbatim}

\begin{center}
\begin{verbatim}
# Function to calculate edge weight based on atomic properties
def calculate_edge_weight(bond):atom1,atom2,bond_type=bond.GetBeginAtom().GetSymbol(),
 bond.GetEndAtom().GetSymbol(),bond.GetBondTypeAsDouble()
atomic_AtomicRadius1=atomicProperties[atom1]["AtomicRadius"]
atomic_AtomicRadius2=atomicProperties[atom2]["AtomicRadius"]
bond_order=int(bond_type)
return (atomicProperties["C"]["AtomicRadius"] * atomicProperties["C"]["AtomicRadius"])/
(bond_order*atomic_AtomicRadius1 *atomic_AtomicRadius2)
\end{verbatim}
\end{center}

\begin{verbatim}
	# List to store weighted graphs
	weighted_graphs = []
	
	# Process each SMILES string
	for smiles in smiles_list:
\end{verbatim}

\begin{verbatim}
	
mol = Chem.MolFromSmiles(smiles)
# Compute vertex weights
vertex_weights = [calculate_vertex_weight(atom.GetSymbol()) for atom in mol.GetAtoms()]
# Compute edge weights
edge_weights=[calculate_edge_weight(bond) for bond in mol.GetBonds()]
# Store the weighted graph
weighted_graphs.append({"SMILES":smiles,"VertexWeights":
	vertex_weights, "EdgeWeights":edge_weights})
\end{verbatim}

\begin{verbatim}
	# Function to calculate the degree of an atom
	def calculate_degree(atom, graph):
	atom_symbol = atom.GetSymbol()
	vertex_weights = graph["VertexWeights"]
	edge_weights = graph["EdgeWeights"]
\end{verbatim}

\begin{verbatim}
 # Calculate the vertex weight of the atom
vertex_weight = vertex_weights[atom.GetIdx()]

# Calculate the sum of edge weights connected to the atom
connected_atoms = [bond.GetOtherAtom(atom) for bond in atom.GetBonds()]
edge_weight_sum = sum(edge_weights[bond.GetIdx()] for bond in atom.GetBonds())
\end{verbatim}

\begin{verbatim}
# Calculate the degree of the atom
degree = vertex_weight + edge_weight_sum

return degree
\end{verbatim}

\begin{verbatim}
# Compute the degree of atoms
degrees = []
for graph in weighted_graphs:
mol = Chem.MolFromSmiles(graph["SMILES"])
atom_degrees = [calculate_degree(atom, graph) for atom in mol.GetAtoms()]
degrees.append(atom_degrees)
\end{verbatim}

\begin{verbatim}
def calculate_SC_index(graph):
atom_degrees = graph["Degrees"]
mol = Chem.MolFromSmiles(graph["SMILES"])
bonds = mol.GetBonds()
SC_index = sum(1/math.sqrt((atom_degrees[bond.GetBeginAtom().GetIdx()]
+atom_degrees[bond.GetEndAtom().GetIdx()]))  for bond in bonds)
return SC_index
\end{verbatim}

\begin{verbatim}
# Compute the SC index of atoms
SC_indices = []
for graph in weighted_graphs:
SC_index = calculate_SC_index(graph)
SC_indices.append(SC_index)  
\end{verbatim}

\begin{verbatim}
# Iterate over the data and populate the table rows
for i, graph in enumerate(weighted_graphs):
\end{verbatim}

\begin{verbatim}
print("SC_index:", SC_indices[i])
print()
\end{verbatim}

\end{appendices}
\end{document}